\documentclass[twocolumn]{aastex631}
\usepackage{mathtools}
\usepackage[frak=esstix]{mathalpha}

\begin{document}

\title{Testing SALT Approximations with Numerical Radiative Transfer Code. II. Thermal and Microturbulent Line Broadening}

\author[0000-0003-4166-2855]{Cody A. Carr}
\affiliation{Center for Cosmology and Computational Astrophysics,
Institute for Advanced Study in Physics, Zhejiang University,
Hangzhou 310058, China}
\affiliation{Institute of Astronomy, School of Physics,
Zhejiang University, Hangzhou 310058, China}
\affiliation{Department of Astronomy, The University of Michigan,
1085 S. University Avenue, West Hall 323,
Ann Arbor, MI 48109, USA}
\correspondingauthor{Cody A. Carr}
\email{codycarr24@gmail.com}

\author[0000-0001-8531-9536]{Renyue Cen}
\affiliation{Center for Cosmology and Computational Astrophysics,
Institute for Advanced Study in Physics, Zhejiang University,
Hangzhou 310058, China}
\affiliation{Institute of Astronomy, School of Physics,
Zhejiang University, Hangzhou 310058, China}
\correspondingauthor{Renyue Cen}
\email{renyuecen@zju.edu.cn}

\author{Leo Michel-Dansac}
\affiliation{Aix Marseille Univ, CNRS, CNES, LAM, Marseille, France}

\author[0000-0002-9136-8876]{Claudia Scarlata}
\affiliation{Minnesota Institute for Astrophysics,
School of Physics and Astronomy, University of Minnesota,
316 Church Street SE, Minneapolis, MN 55455, USA}

\author[0000-0002-6586-4446]{Alaina Henry}
\affiliation{Space Telescope Science Institute,
3700 San Martin Drive, Baltimore, MD 21218, USA}
\affiliation{Center for Astrophysical Sciences,
Department of Physics \& Astronomy, Johns Hopkins University,
Baltimore, MD 21218, USA}



\begin{abstract}
Forward models that connect galactic winds to their predicted spectral-line profiles have proved effective for inferring wind properties in controlled settings, but important limitations remain. In particular, many models rely on the Sobolev approximation to solve the radiative transfer equation and neglect line broadening caused by thermal and turbulent motions within the wind. In the first paper of this series, we demonstrated that neglecting this broadening in Semi-Analytical Line Transfer (SALT) models can bias the recovery of fundamental wind properties from mock observations. Here, we extend the SALT framework to incorporate this motion by solving the radiative transfer equation in the single-scattering limit. We treat re-emission using an escape-probability approach similar to that adopted under the Sobolev approximation, while allowing photons to escape from resonance regions of finite thickness. We validate the model and investigate parameter degeneracies by fitting mock spectra generated with Monte Carlo radiative transfer simulations assuming identical outflow configurations. We identify a degeneracy between the Doppler-broadening parameter and the radial density and velocity profiles: shallower density and velocity gradients can mimic the effects of greater velocity dispersion. Nevertheless, integrated quantities are well recovered. Over the range $13\leq\log(N_{\mathrm{Si}^+}/\mathrm{cm}^{-2})\leq18$, the recovered ionic column densities have a scatter of 0.26 dex and are systematically overestimated by 0.22 dex. Mass-outflow rates evaluated at the terminal wind radius have a scatter of 0.88 dex and are systematically overestimated by 0.51 dex. These results represent substantial improvements over previous versions of the model.
\end{abstract}



\section{Introduction} \label{sec:intro}

Modeling absorption and emission lines in down-the-barrel spectroscopy—in which a galaxy is observed against its own background radiation—provides a means of constraining the physical properties of galactic winds. Forward-modeling approaches (e.g., \citealt{Scarlata2015,Carr2018,Carr2023,Yuan2023,Li2024,Li2026_Model}), which connect physical descriptions of winds to predicted spectral-line profiles, are especially promising. Their ability to capture both the kinematics and physical structure of a wind—including its density and spatial distribution—within a consistent framework makes them particularly valuable for extracting aggregate quantities such as column densities and mass-outflow rates.

However, many of these models have important limitations. Numerical efficiency often comes at the expense of physical realism, whereas greater physical complexity can introduce additional parameter degeneracies \citep{Gronke2015,Li2022}. To accelerate radiative-transfer calculations, many approaches adopt the Sobolev approximation \citep{Lamers1999}. This approximation localizes the interaction of photons with a moving medium to a narrow resonance region and assumes that intrinsic line broadening is negligible relative to the velocity gradient across the flow. It is therefore expected to perform best for low-ionization species that trace relatively cool gas \citep{Zhu2015,Chisholm2016a,Chisholm2017a,Carr2021a,Carr2025_LyC,Huberty2024,Huberty2026,Xu2025}. The approximation can break down for higher-ionization species that trace hotter gas, where thermal broadening may become important, and at high column densities, where absorption in the line wings can no longer be neglected \citep{Li2026_Data}.

In the first paper of this series, we demonstrated that semi-analytical line transfer (SALT) models employing the Sobolev approximation accurately recover the column densities and mass-outflow rates of low-column-density winds, \(\log(N_{\rm ion}/\mathrm{cm}^{-2})\lesssim15\) \citep[hereafter CC23]{Carr2023}. Systematic biases emerge, however, at higher column densities. \citet{Carr2025_FIRE} subsequently found similar biases in column densities inferred from synthetic spectra generated by post-processing cosmological zoom-in simulations from the FIRE-2 suite, demonstrating that the problem persists in more physically realistic environments. Empirical methods also exhibit substantial biases when applied to mock spectra and therefore do not necessarily overcome these limitations \citep{Savage1991,Churchill2015,delaCruz2021,Huberty2024,Jennings2025,Carr2025_FIRE}.
 
In this paper, we extend the SALT model presented in CC23 to account for thermal and turbulent motion in a bi-conical outflow.  While we attempt to preserve the majority of the analytical framework of the original model, we explicitly solve the radiative transport equation without invoking the Sobolev approximation \citep{Chandrasekhar1960}. Although other models (e.g., \citealt{Krumholz2017,Li2024}) also solve the full radiative transport equation, our approach is designed to preserve the analytical simplicity and physical transparency of the original SALT model.\footnote{The new SALT code is publicly available at \href{https://semi-analytic-line-transfer-salt.readthedocs.io}{\texttt{semi-analytic-line-transfer-salt.readthedocs.io}}.}

This paper is organized as follows. In Section~2, we derive the absorption and emission components of the new model for a spherical wind, generalize the formalism to bi-conical outflow geometries, and describe our treatment of multiple resonant and fluorescent scattering. In Section~3, we compare the model predictions directly with those of Monte Carlo radiative-transfer simulations. In Section~4, we test the model’s ability to recover the physical properties of simulated outflows, interpret the results, and examine the model’s limitations. We summarize our conclusions in Section~5.


\section{Modeling}\label{sec:model}


In this section, we adapt the SALT model from CC23 to account for thermal and turbulent motion in the gas frame.  Our goal is to develop a model that accurately reproduces the predictions from numerical scattering experiments at a level of accuracy suitable for data interpretation. Since this paper is part of a series, we refrain from redefining terms relevant to the SALT formalism except where necessary and instead refer the reader to CC23.

\subsection{Absorption}

\begin{figure*}
	\centering
	\includegraphics[width=\textwidth]{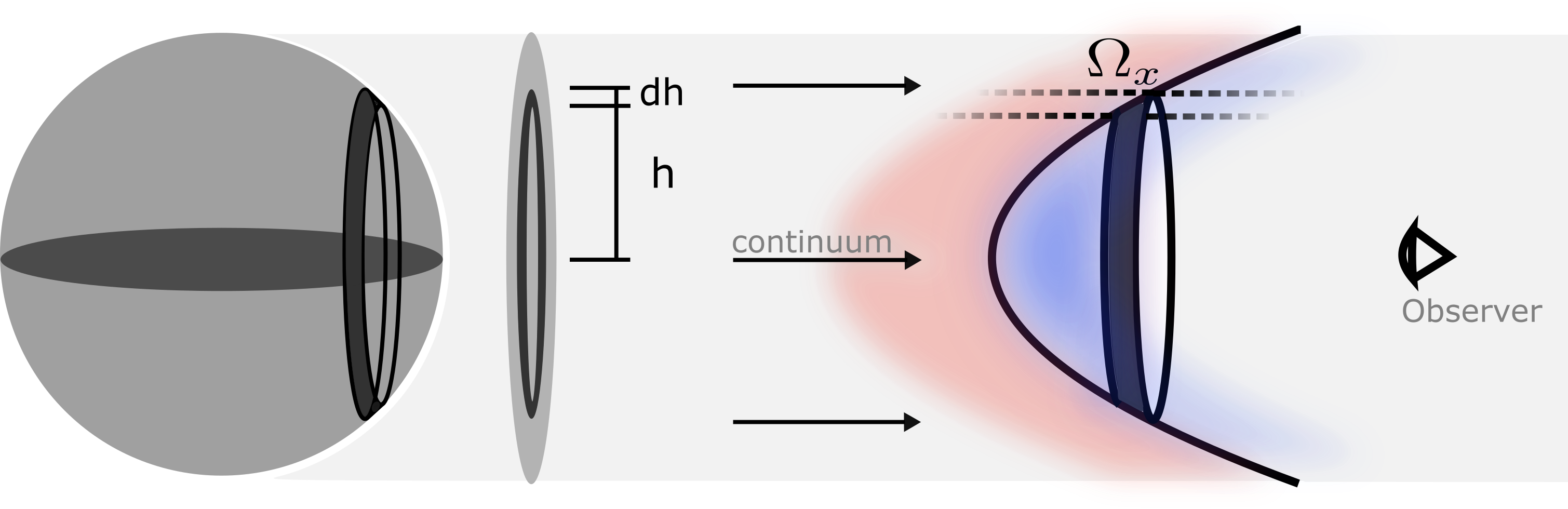}
    	\caption{Schematic of a spherical source emitting radiation isotropically and surrounded by an outflow of material (see Figure~2 of CC23 for further details). Under the Sobolev approximation, a photon of wavelength $\lambda$ is absorbed only within a thin resonance surface, represented by the black structure. When thermal and turbulent motions are included, this surface, $\Omega_x$, broadens into a 3-Dimensional resonance region over which the photon can be absorbed. The red and blue regions indicate gas with projected velocities redward and blueward, respectively, of the Sobolev resonance in the accelerating velocity field adopted by SALT. We must therefore calculate the optical depth along the photon’s path through the wind rather than evaluate the absorption only at the Sobolev resonance surface when accounting for thermal and turbulent motion in the wind.}
	\label{fig:absorption}
\end{figure*} 

\begin{figure*}
	\centering
	\includegraphics[width=\textwidth]{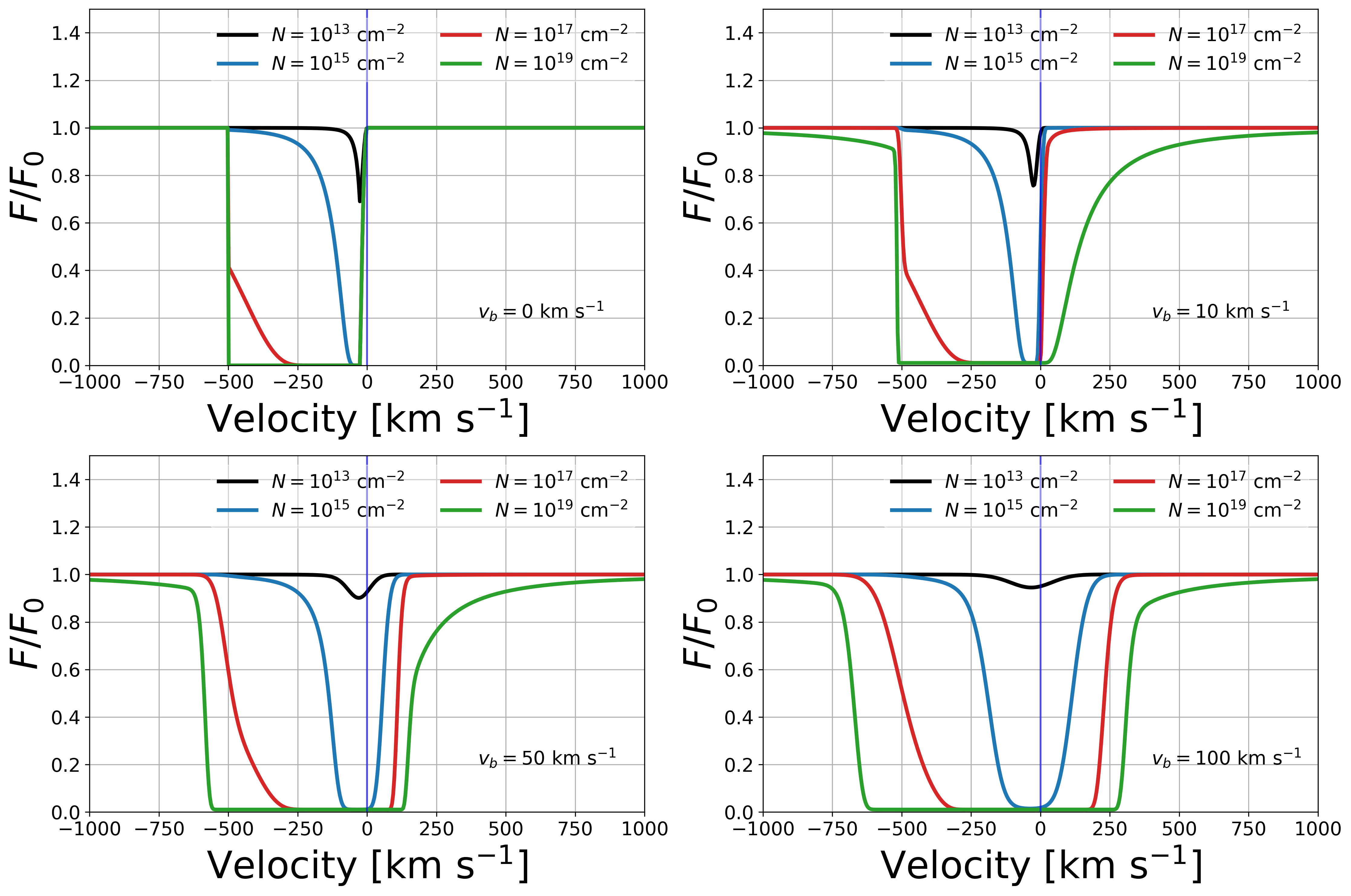}
	\caption{Absorption line profiles for spherical $(\alpha = 90^{\circ},\psi = \rm NA)$ outflows ranging in column density from $10^{13}-10^{19}\ \rm cm^{-2}$ for $v_{b} = 0$ (upper left), 10 (upper right), 50 (lower left), and $100\rm \ km\ s^{-1}$ (lower right).  When $v_b$ increases at $N\leq 10^{17}\ \rm cm^{-2}$, we see a strong absorption feature appear near zero observed velocity which extends further to the red with increasing column density.  This corresponds to the Maxwellian component of the cross section.  This broadening only begins to appear at $v_{\rm b} > 10\ \rm km\ s^{-1}$ which is consistent with the Sobolev approximation.  At high column density, $N\geq 10^{17}\ \rm cm^{-2}$, we begin to see the formation of a shallow tail at highly redshifted and blueshifted velocities.  This feature is due to the Lorentzian component of the cross section.  We see significant deviations from the Sobolev approximation at column density $N> 10^{17}\ \rm cm^{-2}$ and large Doppler parameters  $v_{\rm b} > 10\ \rm km\ s^{-1}$.  The rest of the parameters used in these models are $f_{lu} = 0.277$, $\lambda_{lu} = 1190.42$, $A_{ul} = 6.53\times10^8$, $R_{SF} = 1 \ \rm kpc$, $\gamma = 1.0$, $v_0 = 25 \ \rm{km \ s^{-1}}$, $v_{\infty} = 500 \ \rm{km \ s^{-1}}$, $v_{ap} = 500 \ \rm{km \ s^{-1}}$, $f_c = 1$, $\delta = 3.0$, and $\kappa = 0$.}
	\label{fig:abs_profiles}
\end{figure*} 

We begin with the case of a spherical (Sp) outflow.  In CC23, the absorption profile is derived by focusing on the energy removed from the continuum at a fixed wavelength, $\lambda$, by the corresponding resonant surface, $\Omega_x$.  If we consider random motion in the wind, then the surface $\Omega_x$ will have a width over which photons can be absorbed.  This scenario is illustrated in Figure~\ref{fig:absorption}.

Accordingly, we adapt Equation 6 (CC23) to allow for absorption of the continuum by the full extent of the outflow.  Note that it is now possible for the wind to absorb photons which were previously out of resonance with the wind under the Sobolev approximation.  For this reason, rather than parameterizing Equation 6 of CC23 in terms of the wind’s normalized velocity $y$ and the normalized observed velocity $x$, we use the generic coordinate $h$, measured perpendicular to the $s$-axis, where the $s$-axis is defined to be parallel to the line of sight.  We have
\begin{eqnarray}
    I^{\rm Sp}_{\rm abs, blue}(x)/I_0 &=& \frac{F_{\lambda}}{F_{c,\lambda}} - \frac{2F_{\lambda}}{F_{c,\lambda}} \int_{0}^{1} \nonumber\\ 
    &\times& h \left(1-e^{-\tau(h)}\right)dh, \label{eq:absorption_profile}
\end{eqnarray}
where
\begin{eqnarray}
\tau(h) = \int_{S_{SF}}^{S_{W}} n(s^{\prime},h) \sigma(\nu_{s^{\prime}}) ds^{\prime},
\label{eq:tau}
\end{eqnarray}
is the general expression for the optical depth with cross section $\sigma$, density $n$, and the integral is taken in the direction of the $s$-axis from the edge of the source at $S_{\rm SF}= \sqrt{1-h^2}$ to the edge of the outflow at $S_{\rm W}=\sqrt{y_{\infty}^{2/\gamma}-h^2}$ (see Figure 1, CC23).  We assume a Voigt profile to describe the cross section which in the frame of the wind can be computed for the $l$ to $u$ transition as 

\begin{eqnarray}
    \sigma_{lu}(\nu_{s}) = \frac{\sqrt{\pi}e^2f_{lu}}{m_e c \Delta\nu_{D}}H(a,b),
\end{eqnarray}
where
\begin{eqnarray}
    H(a,b) = \frac{a}{\pi}\int_{-\infty}^{\infty} \frac{e^{-p^2}}{(p-b)^2+a^2} dp,
    \label{eq:Voight}
\end{eqnarray}
$m_e$ is the mass of the election, $c$ is the speed of light, $e$ is the charge of the electron, and $\Delta \nu_D = (v_b/c)\nu_{lu}$ is the Doppler width of Doppler parameter $v_b = \sqrt{v_{\rm th}^2 + v_{\rm turb}^2}$, set by the thermal ($v_{\rm th}$) and turbulent ($v_{\rm turb}$) velocity dispersions, and $\nu_{ul}$ is the frequency of the relevant transition.  $a=(A_{ul})/(4\pi \Delta_{\nu_D})$, $b=(\nu_s-\nu_{lu})/\Delta \nu_D$, and $\nu_s$ is the frequency of the incoming photon in the frame of the wind at location $s$.  In this setting, 
\begin{align}
    \nu_s = \frac{c}{c+v_0(x_s-x)}\nu_{lu},
\end{align}
where $x_s$ is the projection of the velocity field onto the path of the continuum ray,
\begin{align}
     x_s = (s^2+h^2)^{(\gamma-1)/2}s.
\end{align}
The density field can be evaluated along the same path as 
\begin{equation}
    n(s,h) = n_0\left(s^2+h^2\right)^{-\delta/2}.
\end{equation}

For computational convenience, we evaluate Equation~\ref{eq:Voight} using the Faddeeva function,
\begin{eqnarray}
    w(z) &\coloneqq& e^{-z^2}\operatorname{erfc}(-iz)\\
    &=& e^{-z^2}\left (1+\frac{2i}{\sqrt{\pi}}\int_0^ze^{t^2}dt\right),
    \label{eq:fadeeva}
\end{eqnarray}
where $H(a,b) = \mathfrak{Re}{(w)}$, $z = a+bi$, and $\rm{erfc}$ denotes the complementary error function.  In addition, to accelerate calculations, SALT provides the option of approximating $H(a,b)$ using the continued-fraction expansion described in Appendix~A1 of \citet{Smith2015}.


We show absorption profiles for spherical outflows with column densities ranging from $10^{13}$ to $10^{19}\,\mathrm{cm^{-2}}$ and Doppler parameters of 0, 10, 50, and $100\,\mathrm{km\,s^{-1}}$ in Figure~\ref{fig:abs_profiles}.  The $v_{b}=0\ \mathrm{km\,s^{-1}}$ case is solved using the models of CC23, under the Sobolev approximation. For $v_{\rm b}=10\ \mathrm{km\,s^{-1}}$, major deviations from the $v_{b}=0\ \mathrm{km\,s^{-1}}$ case emerge only at $N>10^{17}\ \mathrm{km\,s^{-1}}$.  The winds with larger Doppler parameters ($v_{b}> 10\ \mathrm{km\,s^{-1}}$) show more enhanced redshifted absorption at all column densities $N>10^{13}\ \mathrm{km\,s^{-1}}$.  The differences are marked by enhanced broadening at large optical depths, due to the Maxwellian portion of the cross section.  At column densities $N\sim 10^{19}\ \mathrm{km\,s^{-1}}$, the Lorentzian portion becomes significant, showing absorption at observed velocities greater than the terminal velocity of the wind ($ > 500 \rm \ km\ s^{-1}$) and absorption red ward of the transition ($ < 0 \rm \ km\ s^{-1}$).  The Ly-$\beta$ profiles presented by \citet{Carr2025_Nat} provide examples of spectral lines exhibiting these high-density characteristics.

\begin{figure*}
  \centering
\includegraphics[width=\textwidth]{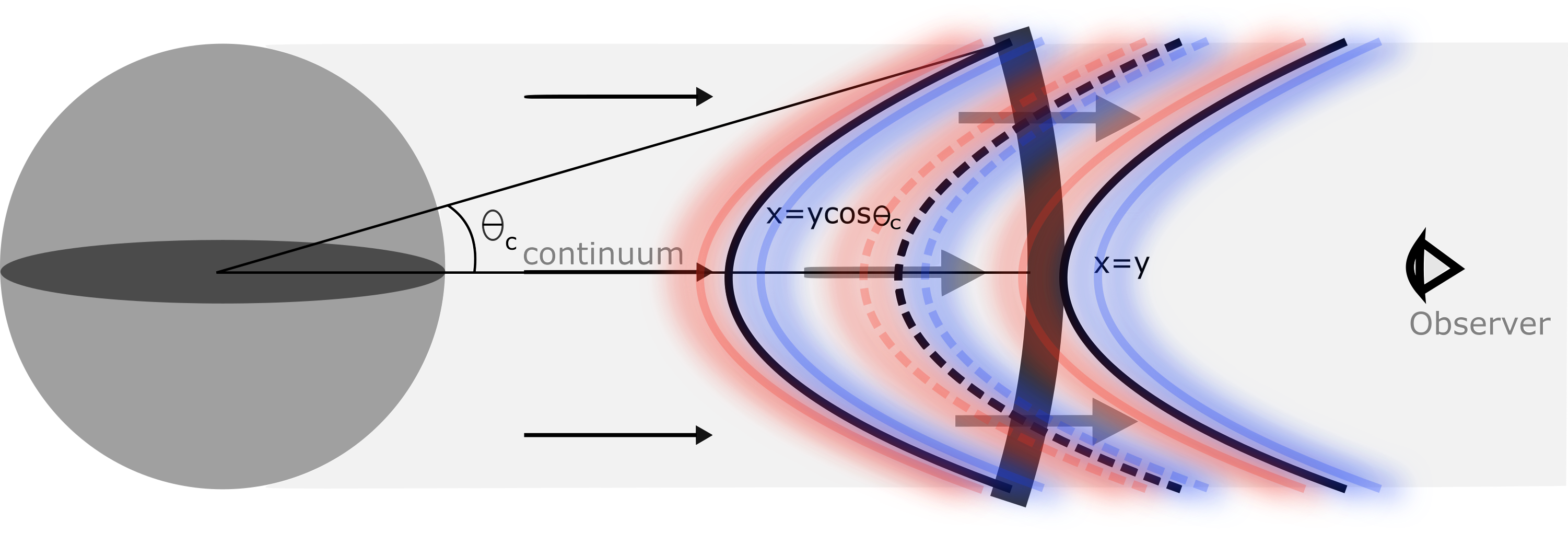}
 \caption{Same base setup as Figure~\ref{fig:absorption}.  For further details, refer to CC23, Figure 4.  We aim to compute the amount of energy absorbed by the shell.  To do this, we first find the amount of energy absorbed where the surfaces $\Omega_x$ intersect the shell.  This does not count for all absorption, however, as turbulence allows for absorption to occur outside of resonance.  As described in the text, we sum together the absorption occurring over all wavelengths in the shell.  After which, we integrate over the different surface to obtain the total energy absorbed by the shell from the continuum, as we continuously deform from one surface to the next in $x$.}
 \label{fig:emission}
\end{figure*} 

\subsection{Emission}

We use the same strategy to compute the emission profile, adapting the derivation in CC23 as needed. Our goal is to compute the total energy absorbed by the outflow within a shell of velocity y and thickness $\Delta y$, and then project this energy into the observed velocity space. In CC23, we first determined the fraction of continuum flux absorbed at resonance by $\Omega_x$ within the shell—that is, the absorption occurring at the intersection of $\Omega_x$ with the shell—and then summed this contribution over all surfaces $\Omega$ that intersect the shell. In CC23, we implemented this procedure by treating $\Omega$ as a continuous family of resonance surfaces (black curves in Figure~\ref{fig:emission}) and smoothly summing the contributions from all absorbing regions as we integrated over $x$.

We follow the same procedure here, but now account for the fact that a photon can be absorbed anywhere within the volume of a shell. Moreover, the radiation incident on the shell may already have been attenuated by
material at smaller radii. The energy absorbed within the shell from photons associated with the resonance surface $\Omega_x$ is therefore given by
\begin{eqnarray}
L^{\rm Sp}_{\text{c,shell},x} &=& \left.\frac{I^{\rm Sp}_{\rm abs, blue}(x)}{I_0}\right|_{y+\Delta y}
- \left.\frac{I^{\rm Sp}_{\rm abs, blue}(x)}{I_0}\right|_{y},
\end{eqnarray}
where the vertical bars denote the radial bounds $y$ and $y+\Delta y$. To include contributions from photons at all wavelengths, we integrate over all observed velocities:
\begin{eqnarray}
L^{\rm Sp}_{\text{c,shell}} = \int_{-\infty}^{\infty} L^{\rm Sp}_{\text{c,shell},x}\, dx.
\label{eq:Lc,shell}
\end{eqnarray}

For a spherical wind, photon absorption by the shell is isotropic. Hence, the fraction of the total flux absorbed by the shell is equal to that of the continuum, i.e.,
\begin{eqnarray}
L^{\rm Sp}_{\text{shell}} = L^{\rm Sp}_{\text{c,shell}}.
\label{eq:Spshell}
\end{eqnarray}

To calculate the red and blue emission components of the line profile, we
substitute our expression for $L^{\rm Sp}_{\mathrm{shell}}$ into Equations~16 and 17
of CC23. If we neglect the effect of thermal and turbulent motion in the gas frame, the normalized blue emission profile for a spherical outflow would be
\begin{equation}
\mathcal{E}^{\rm Sp}_{\mathrm{blue}}(x)
\equiv
\int_{\max(x,1)}^{y_\infty}
\frac{L^{\rm Sp}_{\mathrm{shell}}(y)}{2y}\,dy ,
\label{eq:blue_unbroadened}
\end{equation}
while the corresponding red emission profile would be
\begin{equation}
\mathcal{E}^{\rm Sp}_{\mathrm{red}}(x)
\equiv
\int_{y_1(x)}^{y_\infty}
\frac{L^{\rm Sp}_{\mathrm{shell}}(y)}{2y}\,dy .
\label{eq:red_unbroadened}
\end{equation}
The bounds of integration prevent one from integrating over the source.  Note that $y_1$ can be derived from the following expression:
\begin{eqnarray}
y_1^2(1-y_1^{-2/\gamma}) = x^2.
\label{eq:y1}
\end{eqnarray}

\begin{figure*}
	\centering
	\includegraphics[width=\textwidth]{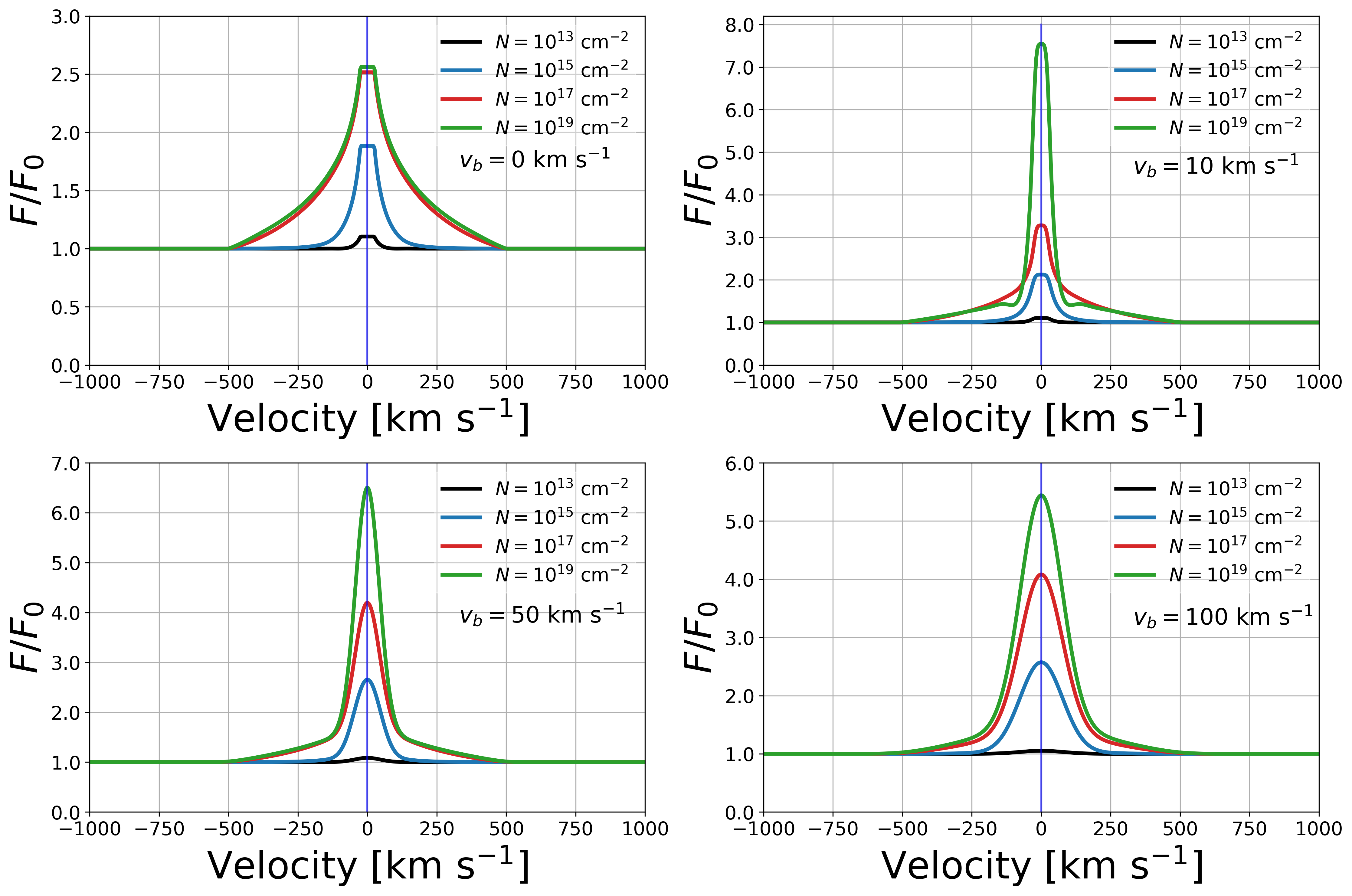}
	\caption{Emission line profiles for spherical ($\alpha = 90^{\circ}, \psi = \mathrm{NA}$) outflows with column densities ranging from $10^{13}$ to $10^{19}\ \mathrm{cm^{-2}}$, for $v_b=0$ (upper left), 10 (upper right), 50 (lower left), and $100\ \mathrm{km\ s^{-1}}$ (lower right). For $v_b=0\ \mathrm{km\ s^{-1}}$, the integrated emission converges with increasing column density because the absorption profile becomes saturated. In contrast, when $v_b>0\ \mathrm{km\ s^{-1}}$, absorption in the wings of the line profile continues to increase with column density, producing progressively stronger emission. Most of the emission is concentrated near zero observed velocity, where the gas density is highest, owing to absorption within the Maxwellian core of the line profile. At the highest column densities ($N\gtrsim10^{17}\ \mathrm{cm^{-2}}$), emission also appears at large observed velocities as photons are absorbed far from line center through the Lorentzian wings. Together, these features resemble multiple kinematic components, although they all originate from the same outflowing gas.  The number of shells used to compute the emission spectrum is 10.  The rest of the parameters used in these models are $f_{lu} = 0.277$, $\lambda_{lu} = 1190.42$, $A_{ul} = 6.53\times10^8$, $R_{SF} = 1 \ \rm kpc$, $\gamma = 1.0$, $v_0 = 25 \ \rm{km \ s^{-1}}$, $v_{\infty} = 500 \ \rm{km \ s^{-1}}$, $v_{ap} = 500 \ \rm{km \ s^{-1}}$, $f_c = 1$, $\delta = 3.0$, and $\kappa = 0$.}
	\label{fig:em_profiles}
\end{figure*}

In reality, the observed frequencies of re-emitted photons are shifted not only by the projected bulk velocity of the wind but also by the thermal and turbulent velocities of the scattering atoms. We assume that these random motions follow an isotropic Maxwellian velocity distribution and neglect the angular dependence of the scattering phase function and atomic recoil \citep[e.g.,][]{Adams1971}. 
We define
\begin{equation}
\sigma_x
=
\frac{\sigma_v}{v_0}
=
\frac{v_b}{\sqrt{2}\,v_0},
\label{eq:sigma_x}
\end{equation}
where $\sigma_{v}=v_b/\sqrt{2}$ is the one-dimensional Gaussian standard deviation.  The normalized Gaussian redistribution kernel is then
\begin{equation}
G(x-x';\sigma_x)
\equiv
\frac{1}{\sqrt{2\pi}\sigma_x}
\exp\left[
-\frac{(x-x')^2}{2\sigma_x^2}
\right].
\label{eq:emission_gaussian}
\end{equation}
The final broadened blue and red emission profiles are obtained by convolving
Equations~\eqref{eq:blue_unbroadened} and
\eqref{eq:red_unbroadened} with this kernel:
\begin{equation}
\frac{I^{\rm Sp}_{\mathrm{em,blue}}(x)}{I_0}
=
\int_{-\infty}^{\infty}
G(x-x';\sigma_x)
\mathcal{E}^{\rm Sp}_{\mathrm{blue}}(x')\,dx'
\label{eq:blue_broadened}
\end{equation}
and
\begin{equation}
\frac{I^{\rm Sp}_{\mathrm{em,red}}(x)}{I_0}
=
\int_{-\infty}^{\infty}
G(x-x';\sigma_x)
\mathcal{E}^{\rm Sp}_{\mathrm{red}}(x')\,dx' .
\label{eq:red_broadened}
\end{equation}

We show emission line profiles for spherical outflows with column densities ranging from $10^{13}$ to $10^{19}\ \mathrm{cm^{-2}}$ and Doppler parameters of 0, 10, 50, and $100\ \mathrm{km\ s^{-1}}$ in Figure~\ref{fig:em_profiles}. Once again, the cases with $v_b=0\ \mathrm{km\ s^{-1}}$ are computed using the Sobolev approximation, as in CC23. In this limit, the integrated emission approaches a limiting value as the absorption profile becomes saturated. In contrast, when $v_b>0\ \mathrm{km\ s^{-1}}$, absorption in the wings of the Voigt profile continues to increase with column density, producing progressively stronger emission. At column densities of $N\gtrsim10^{17}\ \mathrm{cm^{-2}}$, the emission develops distinct low- and high-velocity components that arise from absorption in the Maxwellian core and Lorentzian wings of the absorption cross section, respectively.  The shift of emission from large observed velocities in the $v_b=0\ \mathrm{km\,s^{-1}}$ case toward zero observed velocity in the $v_b=10\ \mathrm{km\,s^{-1}}$ case occurs because most photons are absorbed and re-emitted by the innermost shell of the outflow.

Although the emission-line features originate from a single outflow, simple fitting procedures, such as multicomponent Gaussian decomposition, could easily misinterpret them as distinct kinematic components. Furthermore, comparing the \(v_b=0\ \mathrm{km\,s^{-1}}\) and \(v_b=100\ \mathrm{km\,s^{-1}}\) cases illustrates the substantial effect of thermal and turbulent motions on the line-profile shape. In the \(v_b=100\ \mathrm{km\,s^{-1}}\) case, velocity broadening dominates over the structure imprinted by the density and velocity fields. This effect could be important for studies that use emission-line shapes to distinguish among physical scenarios (e.g., \citealt{Komarova2025}).

\subsection{Bi-conical Wind}  We now generalize the absorption-line profile to that of a bi-conical (BC) outflow with half-opening angle $\alpha$ and orientation angle $\psi$, where $\psi$ is subtended by the cone axis and line of sight. This geometry requires us to account for variations in the path length when integrating the optical depth through the wind. We therefore introduce the azimuthal angle $\varphi=\arctan(l/\xi)$ about the $s$-axis, where $l$ is the Cartesian coordinate perpendicular to both the $s$- and $\xi$-axes (Figure~\ref{fig:abs_geom}). The corresponding azimuthal unit vector is
\begin{equation}
\hat{\boldsymbol{\varphi}}
=
-\sin\varphi\,\hat{\boldsymbol{\xi}}
+\cos\varphi\,\hat{\boldsymbol{l}}.
\end{equation}
In this context, Equation~\ref{eq:absorption_profile} generalizes to  
\begin{eqnarray}
    I^{\rm BC}_{\rm abs, blue}(x)/I_0 &=& \frac{F_{\lambda}}{F_{c,\lambda}} - \frac{1}{\pi}\frac{F_{\lambda}}{F_{c,\lambda}} \int_{0}^{1}\int_0^{2\pi}\nonumber\\ 
    &\times& h \left(1-e^{-\tau(h,\varphi)}\right)d\varphi dh, \label{eq:absorption_profile_bicone}
\end{eqnarray}
where the optical depth,
\begin{eqnarray}
    \tau(\varphi,h) = \int_{S_{1}(\varphi,h)}^{S_{2}(\varphi,h)}n(s^{\prime},h)\sigma({\nu_{s^\prime}})ds,
    \label{eq:tau_bicone}
\end{eqnarray}
is now a function of $h$ and $\varphi$.  $S_{2} - S_{1}$ now defines the path through the bi-cone at height $h$ and rotation angle $\varphi$. We describe how to calculate $S_{2}$ and $S_{1}$ in terms of $h$ and $\varphi$ in the appendix.

To construct the emission profile of a bi-conical outflow, \cite{Carr2018} derived the bi-conical solution by scaling the emission from a spherical shell. This treatment assumes that the radiation incident on each shell is spatially uniform and unaffected by material interior to the shell. These assumptions do not strictly hold for a turbulent bi-conical outflow, in which the incident radiation depends on the angular distribution of the intervening material. Nevertheless, we find that the treatment of \cite{Carr2018} remains highly accurate for the outflows examined in this study (Section~\ref{sec:results}). We therefore adopt a similar approach, replacing Equation~\ref{eq:Lc,shell} with
\begin{equation}
L_{\rm shell}^{\rm BC}(x)=
f_g(x,y)L_{\rm shell}^{\rm Sp},
\label{eq:BCshell}
\end{equation}
where \(f_g\) is the geometric scale factor defined by \cite{Carr2018}. Through its dependence on Equation~\ref{eq:Spshell}, this expression accounts for absorption by the inner shells, likely improving its accuracy. Finally, we compute the normalized emission profiles of a bi-conical outflow by substituting \(L_{\rm shell}^{\rm BC}\) for \(L_{\rm shell}^{\rm Sp}\) in Equations~\ref{eq:blue_broadened} and \ref{eq:red_broadened}:
\begin{equation}
\frac{I^{\rm BC}_{\mathrm{em,blue}}(x)}{I_0}
=
\int_{-\infty}^{\infty}
G(x-x';\sigma_x)
\mathcal{E}^{\rm BC}_{\mathrm{blue}}(x')\,dx'
\label{eq:blue_broadened_BC}
\end{equation}
and
\begin{equation}
\frac{I^{\rm BC}_{\mathrm{em,red}}(x)}{I_0}
=
\int_{-\infty}^{\infty}
G(x-x';\sigma_x)
\mathcal{E}^{\rm BC}_{\mathrm{red}}(x')\,dx' .
\label{eq:red_broadened_BC}
\end{equation}

\subsection{Multiple Scattering}
Thus far, we have considered only the initial absorption of a photon. Re-emitted photons, however, may undergo further absorption and re-emission within the surrounding medium, a process known as radiative trapping. The original SALT model treats these repeated interactions using an escape-probability formalism \citep{Scarlata2015}. Under the Sobolev approximation, the escape probability is
\begin{equation}
\beta(x,y)
= \frac{1-\exp[-\tau_{\mathrm{S}}(x,y)]}
{\tau_{\mathrm{S}}(x,y)},
\label{eq:esc_prob}
\end{equation}
where the Sobolev optical depth \citep{Lamers1999} is given by
\begin{eqnarray}
\tau_S(x,y) &=& \frac{\tau_0}{1+(\gamma-1)\left(x/y\right)^2}y^{(1-\gamma-\delta)/\gamma},
\label{tau_s}
\end{eqnarray}
and
\begin{eqnarray}
\tau_0 = \frac{\pi e^2}{mc} f_{lu}\lambda_{lu} n_0 \frac{R_{SF}}{v_0}.
\label{tau0}
\end{eqnarray}
$R_{\rm SF}$ is the radius of the source.  Note that Equation~\ref{tau0} neglects stimulated emission (see CC23).

Following resonant re-emission, a photon escapes the local resonance region with probability $\beta$ and is reabsorbed with probability $1-\beta$. Depending on the ion and transition, an absorbed photon is re-emitted through the resonant channel with probability $p_R$ or through the fluorescent channel with probability $p_F$, where $p_R+p_F=1$ (see \citealt{Scarlata2015} for further details). The probability that the photon undergoes another absorption–re-emission cycle is therefore $p_R(1-\beta)$. Assuming that fluorescently emitted photons escape the surrounding medium without further interaction, and summing over infinitely many cycles, the fraction of photons that ultimately escape through the resonant channel is
\begin{equation}
F_R = \frac{\beta p_R}{1-p_R(1-\beta)},
\label{eq:resonant_escape_fraction}
\end{equation}
while the fraction that escapes through the fluorescent channel is
\begin{equation}
F_F = \frac{p_F}{1-p_R(1-\beta)}.
\label{eq:fluorescent_escape_fraction}
\end{equation}

This treatment assumes that radiative interactions are localized. Because of the velocity gradient, a re-emitted photon is rapidly Doppler-shifted out of resonance with the surrounding gas, allowing it to escape without further interaction. When thermal and turbulent gas motions are included, however, the extent of the resonance region depends on the velocity gradient, Doppler broadening, and the density of the surrounding medium. Moreover, a photon emitted in one transition may be absorbed by a neighboring transition if their opacity profiles overlap (e.g., Mg II doublet \citealt{Prochaska2011}).  Accounting for these effects requires a more complete treatment of nonlocal line transfer \citep[e.g.,][]{Rybicki1985}.

We find, however, that the original escape-probability treatment remains a good approximation over the range of outflow properties considered here (Section~\ref{sec:results}). We therefore retain the formalism of \citet{Scarlata2015}, but modify it to account for absorption by neighboring transitions, an effect that becomes increasingly important with column density. After calculating the resonant and fluorescent emission produced by each shell using the local escape-probability formalism, we attenuate this emission by the optical depth of all relevant transitions encountered along the photon trajectory between the emitting shell and the outer wind boundary. We calculate this optical depth using the approach described by Equations~\ref{eq:tau}--\ref{eq:fadeeva}.

For a photon emitted from a shell at the line-of-sight position
$S_{\rm shell}=x y^{(1-\gamma)/\gamma}$, the optical depth due to overlapping transitions is
\begin{equation}
\tau^{\mathrm{blend}}_{\rm R/F}(x,y)
=
\sum_j
\int_{S_{\rm shell}(x,y)}^{S_W}
n(s';x,y)\,
\sigma_j\!\left(\nu_{j,s'}\right)
\,ds',
\label{eq:blend_tau}
\end{equation}
where
\begin{equation}
n(s';x,y)
=
n_0
\left[
s'^2+y^{2/\gamma}
-x^2y^{2(1-\gamma)/\gamma}
\right]^{-\delta/2}
\end{equation}
is the density along the photon trajectory and
\begin{equation}
S_W
=
\left[
y_\infty^{2/\gamma}
-y^{2/\gamma}
+x^2y^{2(1-\gamma)/\gamma}
\right]^{1/2}
\end{equation}
is the position at which the trajectory exits the wind. The sum extends over all transitions that overlap the emitted line.

This calculation assumes a spherical wind geometry; adopting the full biconical geometry would require an additional geometric integration. We find that the spherical approximation performs well over the range of outflow configurations considered here (Section~\ref{sec:results}). The projected velocity at position $s'$, expressed in terms of $x$ and $y$, is
\begin{equation}
x_{s'}
=
s'
\left[
s'^2+y^{2/\gamma}
-x^2y^{2(1-\gamma)/\gamma}
\right]^{(\gamma-1)/2},
\end{equation}
which is used to evaluate $\sigma_j(\nu_{j,s'})$. We then attenuate the emission from each shell by the corresponding transmission,
\begin{equation}
T_{\mathrm{R/F}}^{\mathrm{blend}}(x,y)
\equiv
\exp\!\left[-\tau_{\mathrm{R/F}}^{\mathrm{blend}}(x,y)\right].
\label{eq:blend_transmission}
\end{equation}

Finally, we integrate the attenuated contributions from all emitting shells to obtain the blue- and red-side emission profiles:
\begin{align}
\mathcal{E}_{\mathrm{blue}}^{\mathrm{BC,blend}}(x)
={}&
\int_{\max(x,1)}^{y_\infty}
T_{\mathrm{R}}^{\mathrm{blend}}(x,y)\,
F_{\mathrm{R}}(y)\,
\frac{L_{\mathrm{shell}}^{\mathrm{BC}}(x,y)}{2y}
\,dy
\nonumber\\
&+
\int_{\max(x,1)}^{y_\infty}
T_{\mathrm{F}}^{\mathrm{blend}}(x,y)\,
F_{\mathrm{F}}(y)\,
\frac{L_{\mathrm{shell}}^{\mathrm{BC}}(x,y)}{2y}
\,dy,
\label{eq:blue_unbroadened_blended}
\end{align}
and
\begin{align}
\mathcal{E}_{\mathrm{red}}^{\mathrm{BC,blend}}(x)
={}&
\int_{y_1}^{y_\infty}
T_{\mathrm{R}}^{\mathrm{blend}}(x,y)\,
F_{\mathrm{R}}(y)\,
\frac{L_{\mathrm{shell}}^{\mathrm{BC}}(x,y)}{2y}
\,dy
\nonumber\\
&+
\int_{y_1}^{y_\infty}
T_{\mathrm{F}}^{\mathrm{blend}}(x,y)\,
F_{\mathrm{F}}(y)\,
\frac{L_{\mathrm{shell}}^{\mathrm{BC}}(x,y)}{2y}
\,dy.
\label{eq:red_unbroadened_blended}
\end{align}

\subsection{Complete Model}

In the complete model, we account for dust attenuation by multiplying the integrands in Equations~\ref{eq:blue_unbroadened_blended} and \ref{eq:red_unbroadened_blended} by \(e^{-\tau_{\rm dust}}\), following \citet{Carr2021a}. Similarly, we multiply the integrands by \(\Theta_{\rm AP}\) to account for a finite circular observing aperture, following \citet{Carr2018}. The original SALT model parameterizes the porosity of the outflow through the scale factor $f_c$, which is typically assumed to be independent of radius. We retain this assumption here. In future work, we will consider radially varying prescriptions for $f_c$ and investigate the relationship between the two-dimensional covering factor and the volume filling factor of the wind. Combining these effects, the complete line profile is given by

\begin{widetext}
\begin{equation}
\begin{aligned}
&\frac{
 I_{\mathrm{abs,blue}}(x)
 + I_{\mathrm{em,blue}}(x)
 + I_{\mathrm{em,red}}(x)
}{I_0}=
\\
{}&
\frac{F(x)}{F_{c}(x)} - \frac{1}{\pi}\frac{F(x)}{F_{c}(x)} \int_{0}^{1}dh\int_0^{2\pi}d\varphi\,
    h \left(1-e^{-\tau(h,\varphi)}\right)
\\
&+
\int_{-\infty}^{\infty}
dx'G(x-x';\sigma_x)
\int_{\max(x',1)}^{y_\infty}dy
T_{\mathrm{R}}^{\mathrm{blend}}(x',y)\,
F_{\mathrm{R}}(x',y)\,f_c\, e^{-\tau_{\rm dust}(x',y)}\,\Theta_{\rm AP}(x',y)\,f_g(x',y)\,
\frac{L_{\mathrm{shell}}^{\rm Sp}(y)}{2y}
\\
&+
\int_{-\infty}^{\infty}
dx'G(x-x';\sigma_x)
\int_{\max(x',1)}^{y_\infty}dy
T_{\mathrm{F}}^{\mathrm{blend}}(x',y)\,
F_{\mathrm{F}}(x',y)\,f_c\, e^{-\tau_{\rm dust}(x',y)}\,\Theta_{\rm AP}(x',y)\,f_g(x',y)\,
\frac{L_{\mathrm{shell}}^{\rm Sp}(y)}{2y}
\\
&+
\int_{-\infty}^{\infty}
dx'G(x-x';\sigma_x)
\int_{y_1(x')}^{y_\infty}dy
T_{\mathrm{R}}^{\mathrm{blend}}(x',y)\,
F_{\mathrm{R}}(x',y)\,f_c\, e^{-\tau_{\rm dust}(x',y)}\,\Theta_{\rm AP}(x',y)\,f_g(x',y)\,
\frac{L_{\mathrm{shell}}^{\rm Sp}(y)}{2y}
\\
&+
\int_{-\infty}^{\infty}
dx'G(x-x';\sigma_x)
\int_{y_1(x')}^{y_\infty}dy
T_{\mathrm{F}}^{\mathrm{blend}}(x',y)\,
F_{\mathrm{F}}(x',y)\,f_c\, e^{-\tau_{\rm dust}(x',y)}\,\Theta_{\rm AP}(x',y)\,f_g(x',y)\,
\frac{L_{\mathrm{shell}}^{\rm Sp}(y)}{2y}.
\end{aligned}
\label{eq:complete}
\end{equation}
\end{widetext}

\section{Comparison with Simulations}
Following CC23, we compare the spectral-line predictions of SALT with those of RAdiation SCattering in Astrophysical Simulations (RASCAS; \citealt{Michel-Dansac2020}). RASCAS is a Monte Carlo radiative transfer code designed to model the resonant and fluorescent scattering of photons through simulation outputs in post-processing \citep{Kimm2022,Rosdahl2022,Katz2022,Katz2023,Blaizot2023,Farcy2025}. RASCAS also includes an idealized-model framework that allows users to construct arbitrary density and velocity fields on a computational grid and propagate photons through them. In CC23, we used this framework to compare RASCAS and SALT predictions for identical outflow configurations. We also developed an adaptive mesh refinement (AMR) procedure that resolves both the outflow boundaries and gradients in the density and velocity fields, which we employ here. For all models considered in this work, we adopt a maximum refinement level of $l_{\rm max}=10$, corresponding to a maximum effective linear resolution of $2^{10}$ cells. Cells are refined when the density or velocity variations exceed their respective refinement thresholds, $\epsilon_{\rho}=0.02$ and $\epsilon_v=0.02$, or when additional refinement is required to resolve the Doppler width $v_b$.

Because RASCAS models multiple scattering while accounting for the full line profile, it provides a means of investigating the effects of thermal and microturbulent motions within the outflow. In CC23, we used RASCAS to test the recovery of outflow properties from synthetic turbulent-wind spectra using the original SALT formalism. We repeat these recovery tests using the updated SALT model in the following section. First, we directly compare the RASCAS and SALT predictions for turbulent biconical outflows.

\begin{figure*}
	\centering
	\includegraphics[width=\textwidth]{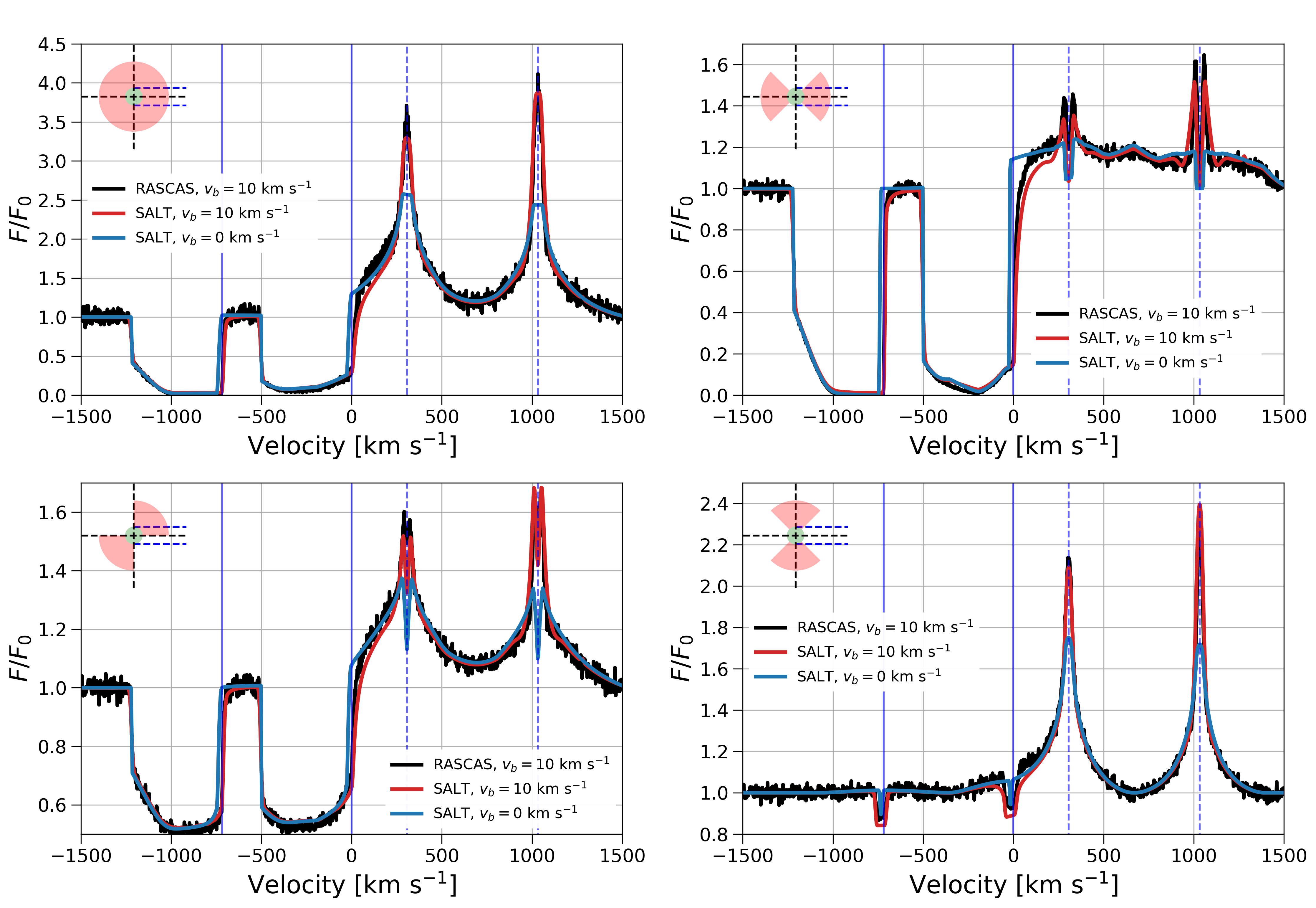}
	\caption{Comparison of RASCAS (red) and SALT (black) predictions for the Si~{\sc ii} $\lambda\lambda1190,1193$ line profiles of outflows viewed at different orientations. From left to right and top to bottom, the panels show a spherical outflow ($\alpha = 90^{\circ},\psi = \ \rm NA$); a bi-conical outflow with its axis parallel to the line of sight ($\alpha = 45^{\circ},\psi = \ 0^{\circ}$); the same outflow viewed at an intermediate inclination ($\alpha = 45^{\circ},\psi = \ 45^{\circ}$); and the outflow viewed with its axis perpendicular to the line of sight ($\alpha = 45^{\circ},\psi = \ 90^{\circ}$). The models adopt a Doppler parameter of $v_b=10\ {\rm km\ s^{-1}}$ and an ionic column density of $\log(N_{\rm Si^+}/{\rm cm}^{-2})=17$. The light-blue curves show predictions from the original SALT prescription of CC23, which assumes $v_b=0\ {\rm km\ s^{-1}}$ and adopts the Sobolev approximation. Including turbulent broadening substantially improves the agreement between SALT and RASCAS at high column densities.  In particular, absorption extending redward of the Sobolev solution enhances the emission profiles.  This improvement is evident for every geometry, including orientations in which emission from the two sides of the bi-cone produces a double-peaked profile.  10 shells are used to compute the emission spectrum in each model.  The remaining model parameters are $R_{SF} = 1 \ \rm kpc$, $\gamma = 1.0$, $v_0 = 25 \ \rm{km \ s^{-1}}$, $v_{\infty} = 500 \ \rm{km \ s^{-1}}$, $v_{ap} = 500 \ \rm{km \ s^{-1}}$, $f_c = 1$, $\delta = 3.0$, and $\kappa = 0$.}
	\label{fig:complete_model}
\end{figure*}

Figure~\ref{fig:complete_model} presents the Si~{\sc ii} $\lambda\lambda1190,1193$ line profiles predicted for biconical outflows with a relatively high ionic column density of $\log(N_{\mathrm{Si}^+}/\mathrm{cm}^{-2})=17$. We compare RASCAS with both the updated SALT model and the original SALT prescription, which adopts the Sobolev approximation and corresponds here to $v_b=0~\mathrm{km\,s^{-1}}$. The original and updated SALT predictions differ substantially at low observed velocities, where the bulk wind velocity is comparable to the Doppler parameter, but converge at high observed velocities, as expected. Overall, the updated SALT model agrees substantially better with the RASCAS simulations than does the original prescription.

\section{Results and Discussion}
\label{sec:results}

In CC23, we evaluated the original SALT model under the Sobolev approximation by applying it, together with a Bayesian Markov chain Monte Carlo fitting procedure \citep{Carr2021a}, to synthetic Si~{\sc ii} $\lambda\lambda1190,1193$ spectra of bi-conical outflows generated with RASCAS. Treating the RASCAS predictions as the ground truth, we investigated whether the line profiles and underlying outflow properties could be recovered accurately when thermal and turbulent line broadening were neglected. We found that the ionic column density and derived quantities, including the mass-outflow rate, were frequently overestimated at high column densities, $\log(N_{\rm Si^+}/\mathrm{cm}^{-2})\gtrsim15$. This trend is consistent with the results presented in Section~\ref{sec:model}, which demonstrate that the effects of thermal and turbulent line broadening increase with column density. \citet{Carr2025_FIRE} subsequently identified a similar bias in column densities recovered from SALT fits to mock spectra generated from the more physically realistic FIRE-2 hydrodynamic zoom-in simulations of galaxy formation. Moreover, empirical methods and partial-covering models have also struggled to recover the true column densities of simulated outflows \citep{Churchill2015,delaCruz2021,Huberty2024,Jennings2025,Carr2025_FIRE}. Incorporating the radiative transfer effects of thermal and turbulent motions into the SALT formalism—and determining whether doing so mitigates these biases—is therefore a primary motivation for the present work.

In this section, we repeat and extend the recovery tests of CC23 to evaluate the updated SALT model. We generate mock spectra with RASCAS using the same wind configurations assumed by SALT and analyze them in the same manner as observations. We use these tests to quantify the accuracy of the recovered wind parameters, identify parameter degeneracies, and determine whether the additional parameter space introduced by the updated model is warranted.

\subsection{Tests against mock spectra}

The mock spectra assembled with RASCAS were drawn from the same parameter space as in CC23, except that we now sample $v_b$ uniformly over $3 < v_b/\mathrm{km\,s^{-1}} < 30$, whereas $v_b = 10\ \rm km\ s^{-1}$ was fixed in CC23.  We generate 150 mock spectra and retain 146, excluding four cases that were so saturated that absorption extended beyond the boundary of the experiment (1188-1200 \AA). This occurs only for the combination of very fast outflows ($>1000\ {\rm km\ s^{-1}}$) and high ionic column densities ($\log{N_{\mathrm{Si}^{+}}/{\rm cm}^{-2}}>18$). Because such high column densities are expected to be rare \citep{Huberty2024,Jennings2025,Carr2025_FIRE}, excluding these cases does not materially affect our conclusions.  We use the same priors and initial conditions as CC23 for SALT during model fitting over the old parameter space.  For the new parameter, $v_b$, we adopt a uniform prior over
$3 < v_b/\mathrm{km\,s^{-1}} < 30$ \citep{Chen2023} and initially draw from the same range.  In addition, we assume 10 shells in each model to compute the emission components of the spectra.  

\begin{figure*}
	\centering
	\includegraphics[width=\textwidth]{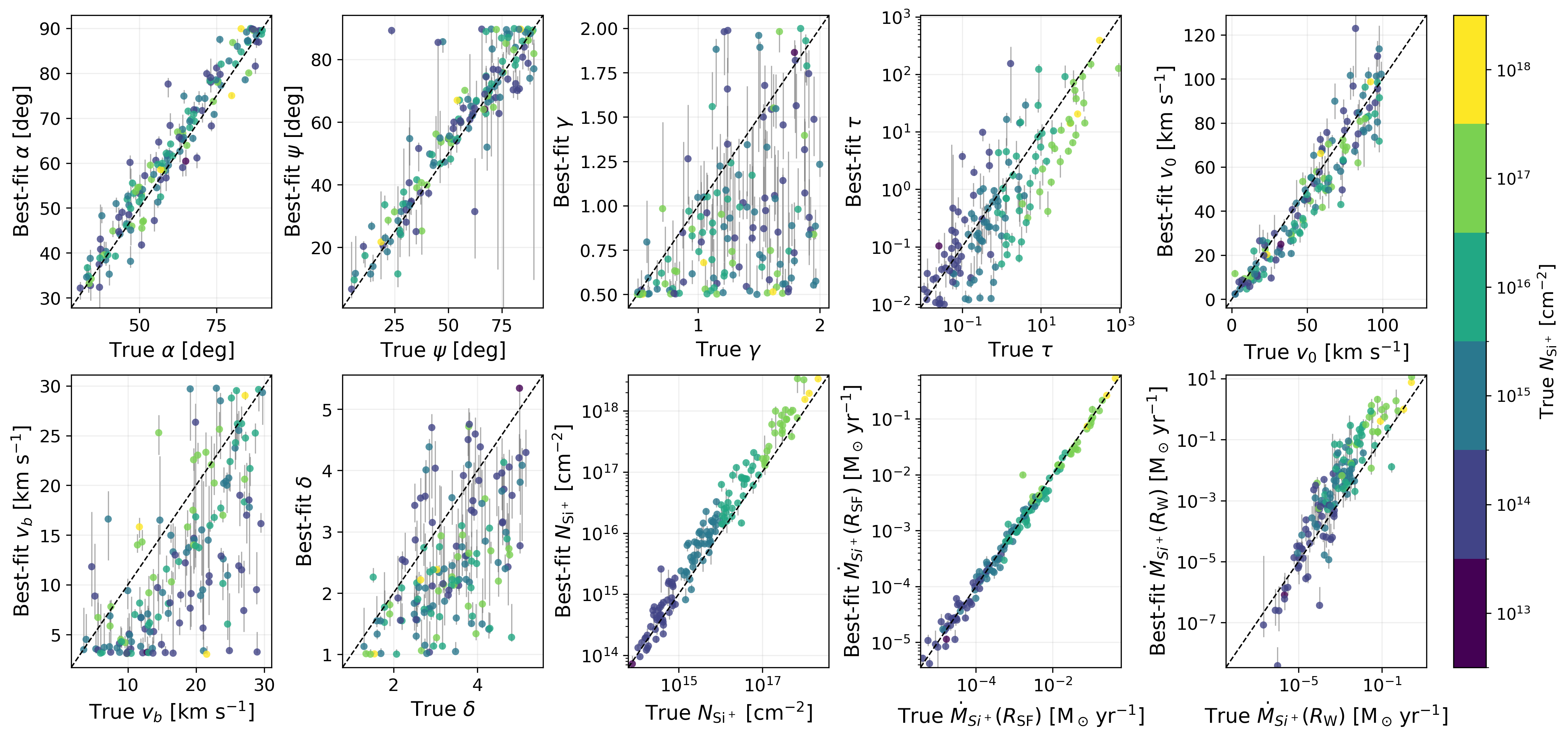}
	\caption{Recovery of SALT parameters from synthetic RASCAS
Si~{\sc ii} $\lambda\lambda1190,1193$ line profiles of turbulent bi-conical outflows. From left to right and top to bottom, the panels show the opening angle ($\alpha$), orientation angle ($\psi$), velocity power-law index ($\gamma$), optical-depth parameter normalized by $\lambda_{lu}f_{lu}$ ($\tau$), launch velocity ($v_0$), Doppler parameter ($v_b$), density power-law index ($\delta$), ionic column density ($N_{\mathrm{Si}^+}$), and Si$^+$ mass-outflow rates evaluated at $R_{\mathrm{SF}}$ and $R_{\mathrm{W}}$, respectively [$\dot{M}_{\mathrm{Si}^+}(R_{\mathrm{SF}})$ and
$\dot{M}_{\mathrm{Si}^+}(R_{\mathrm{W}})$]. The dashed line denotes perfect recovery. Error bars give the median absolute deviations below and above the best-fit value, calculated from the corresponding posterior chain; the uncertainties in the derived quantities are calculated from posterior chains constructed by evaluating each quantity for every joint posterior sample.  Individual points are colored according to column density as indicated by the color bar.  Relative to the original Sobolev SALT model (Figure~17 of CC23), the turbulent model substantially improves the recovery of $v_0$, $\tau$, and $N_{\mathrm{Si}^+}$. However, the recovered $\gamma$ and $\delta$ values now show systematic offsets that were absent or less apparent in the original SALT model.}    
	\label{fig:scatter}
\end{figure*}

As in CC23, we assume normally distributed measurement uncertainties with variances representative of COS observations. We simulate observational noise by drawing the flux in each spectral pixel from a normal distribution centered on the predicted flux, with a standard deviation equal to the adopted measurement uncertainty. The spectra are smoothed to a velocity resolution of $20\ \mathrm{km\,s^{-1}}$ and sampled with three pixels per resolution element.

\begin{figure*}
	\centering
	\includegraphics[width=\textwidth]{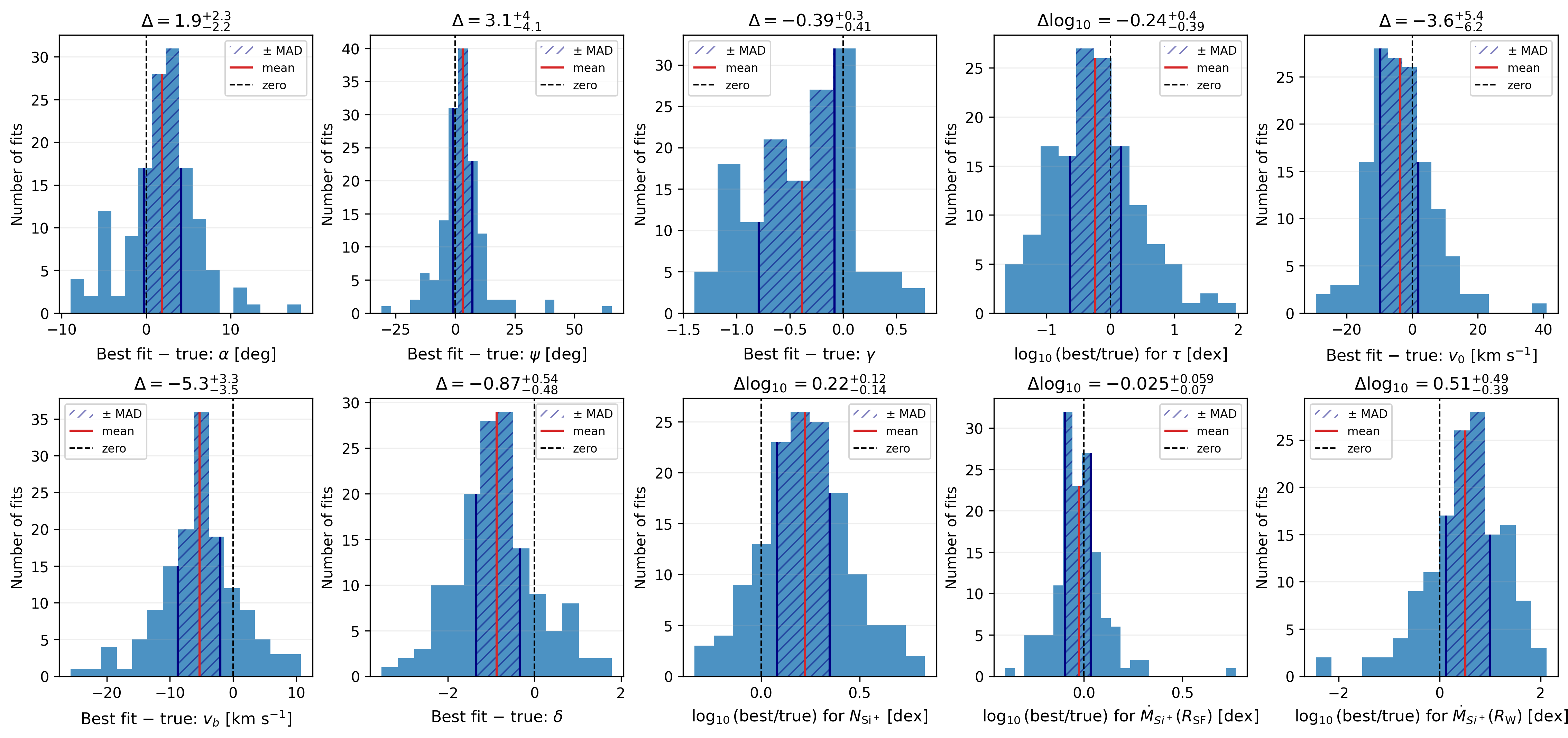}
	\caption{Distributions of the differences between parameters recovered by fitting the SALT model and their true input values for synthetic RASCAS Si~{\sc ii} $\lambda\lambda1190,1193$ line profiles of turbulent bi-conical outflows. From left to right and top to bottom, the panels show the opening angle ($\alpha$), orientation angle ($\psi$), velocity power-law index ($\gamma$), optical-depth parameter normalized by $\lambda_{lu}f_{lu}$ ($\tau$), launch velocity ($v_0$), Doppler parameter ($v_b$), density power-law index ($\delta$), ionic column density ($N_{\mathrm{Si}^+}$), and Si$^+$ mass-outflow rates evaluated at $R_{\mathrm{SF}}$ and $R_{\mathrm{W}}$, respectively [$\dot{M}_{\mathrm{Si}^+}(R_{\mathrm{SF}})$ and $\dot{M}_{\mathrm{Si}^+}(R_{\mathrm{W}})$]. Differences are expressed as best-fit minus true values for linearly plotted parameters and as $\log_{10} (\mathrm{best}/\mathrm{true})$ for logarithmically plotted parameters. The red line marks the mean of each distribution, and the dashed black line marks zero. The navy hatched region spans the asymmetric median absolute deviations below and above the mean. Systematic offsets are evident in the recovery of $\gamma$, $v_b$, $\delta$, and $\dot{M}_{\mathrm{Si}^+}(R_{\mathrm{SF}})$.}
	\label{fig:pdfs}
\end{figure*}

\begin{figure*}
	\centering
	\includegraphics[width=\textwidth]{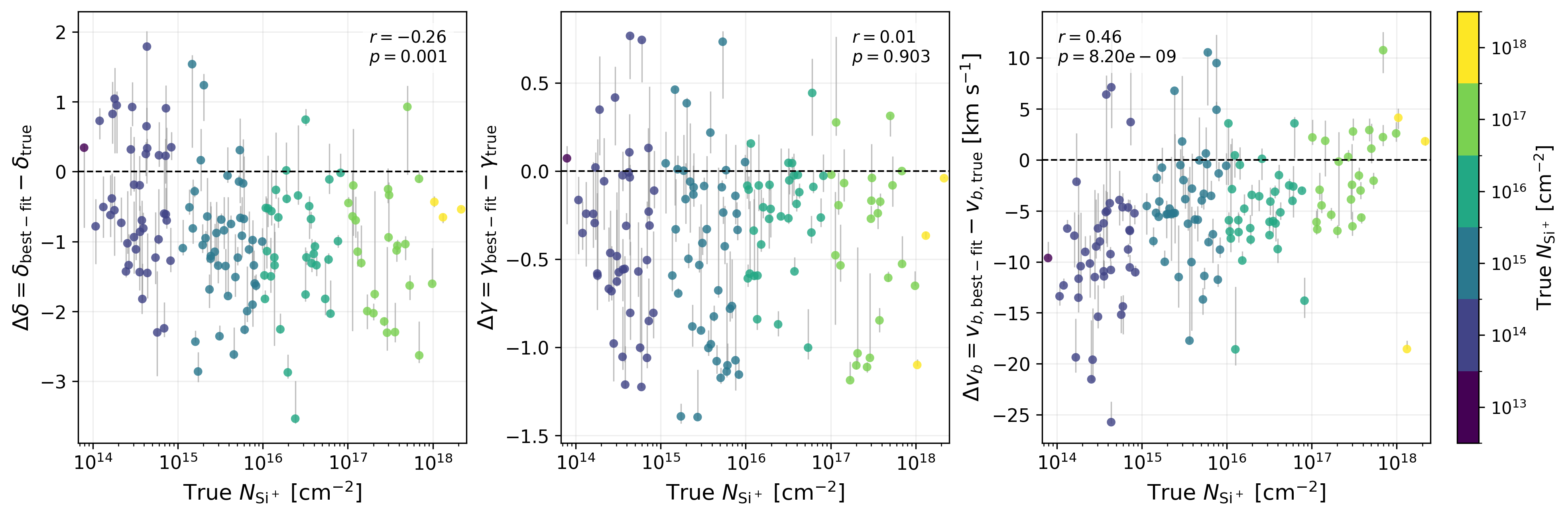}
	\caption{Differences between the SALT-recovered parameters and their true input values for synthetic RASCAS Si~{\sc ii} $\lambda\lambda1190,1193$ line profiles of turbulent biconical outflows, shown as a function of column density. From left to right, the panels present the residuals in $\delta$, $\gamma$, and $v_b$, defined as the best-fit value minus the true value. Points are colored by column density, as indicated by the color bar. Error bars represent the median absolute deviations below and above the best-fit value, calculated from the corresponding posterior distributions. The Spearman rank correlation coefficient and associated $p$-value are reported in each panel. We find a weak negative correlation between the residuals in $\delta$ and $N_{\rm Si^+}$, no significant correlation for $\gamma$, and a moderate positive correlation for $v_b$.}
	\label{fig:diff_vs_cd}
\end{figure*}

Figure~\ref{fig:scatter} shows scatter plots comparing the best-fit SALT parameters with their true input values, colored according to column density, and Figure~\ref{fig:pdfs} shows the corresponding distributions of the differences
between the recovered and true values. We show the opening angle ($\alpha$), orientation angle ($\psi$), velocity power-law index ($\gamma$), optical-depth parameter normalized by $\lambda_{lu}f_{lu}$ ($\tau$), launch velocity ($v_0$), Doppler parameter ($v_b$), and density power-law index ($\delta$). We also consider three derived quantities: the Si$^+$ ionic column density and the Si$^+$ mass-outflow rates evaluated at $R_{\mathrm{SF}}$ and $R_{\mathrm{W}}$, $\dot{M}_{Si^{+}}(R_{\rm SF})$ and $\dot{M}_{Si^{+}}(R_{\rm W})$, respectively.  Written explicitly in terms of the SALT parameters, the Si$^+$ column density is
\begin{equation}
N_{\mathrm{Si}^+} =
\begin{cases}
\displaystyle
\frac{n_{\mathrm{Si}^+,0}R_{\mathrm{SF}}}{\gamma}
\ln\left(\frac{v_\infty}{v_0}\right),
& \delta=1, \\[1em]
\displaystyle
\frac{n_{\mathrm{Si}^+,0}R_{\mathrm{SF}}}{1-\delta}
\left[
\left(\frac{v_\infty}{v_0}\right)^{(1-\delta)/\gamma}-1
\right],
& \delta\ne1,
\end{cases}
\label{eq:N_SiII}
\end{equation}
and the Si$^+$ mass-outflow rate at radius $r$ is
\begin{eqnarray}
\dot{M}_{\rm Si^{+}}(v) &=& 4\pi(1-\cos{\alpha)}R_{\rm SF}^2 \nonumber\\
&&\times v_0m_{\rm Si^{+}}n_{\rm Si^{+},0} \left( \frac{r}{R_{\rm SF}} \right)^{2+\gamma - \delta},
\label{eq:MOR}
\end{eqnarray}
where $f_c$ is the porosity or fraction of the bi-conical shell covered by material. Note that we include $f_c$ in Equation~\ref{eq:MOR} for completeness, and assume $f_c=1$ in all simulations.  The terminal radius can be computed as
\begin{equation}
R_{\mathrm{W}}
=
R_{\mathrm{SF}}
\left(\frac{v_\infty}{v_0}\right)^{1/\gamma}.
\label{eq:Rw}
\end{equation}

Among the directly fitted SALT parameters, $\alpha$, $\psi$, $\tau$, and $v_0$ are generally well recovered. In particular, the recovery of $v_0$ and $\tau$ is substantially improved relative to the original SALT model under the Sobolev approximation (see Figures~17 and 18 of CC23). The geometric parameters, $\alpha$ and $\psi$, are especially tightly constrained, with the largest discrepancies occurring at the lowest column densities, where weaker spectral features are more easily obscured by noise. By contrast, the recovered values of $v_b$, $\gamma$, and $\delta$ exhibit systematic offsets.  $\gamma$ and $\delta$ were largely unconstrained in CC23.

The offsets in $v_b$, $\gamma$, and $\delta$ suggest a degeneracy between thermal or turbulent line broadening and the radial structure of the wind. This degeneracy is expected because absorption produced by thermal and turbulent broadening is strongest at low observed velocities, where the bulk wind speed is comparable to $v_b$ and the density is highest in our adopted density field. A combination of shallow velocity and density gradients can reproduce similar absorption.  Notice that decreasing $\delta$ enhances absorption over a broad range of observed velocities (see the left panel of Figure~12 in CC23). Decreasing $\gamma$, however, causes the wind to accelerate more gradually, shifting gas at high velocities to larger radii and therefore lower densities. This suppresses absorption at high observed velocities (see the right panel of Figure~12 in CC23). Consequently, simultaneous decreases in $\gamma$ and $\delta$ can mimic the spectral effects of thermal or turbulent broadening in an accelerating wind.

In Figure~\ref{fig:diff_vs_cd}, we examine how the differences between the SALT-recovered parameters and the true values used to generate the RASCAS spectra vary with column density for $\delta$, $\gamma$, and $v_b$. We calculate the Spearman rank correlation coefficient for each parameter. We find a weak negative correlation between $\Delta\delta$ and $N_{\rm Si^+}$ ($r_s=-0.26$, $p=0.001$), no significant correlation for $\Delta\gamma$ ($r_s=0.01$, $p=0.903$), and a moderate positive correlation between $\Delta v_b$ and $N_{\rm Si^+}$ ($r_s=0.46$, $p=8.20\times10^{-9}$). These trends suggest a trade-off between $\delta$ and $v_b$: at low column densities, SALT underestimates $v_b$ more strongly and $\delta$ less severely, whereas at higher column densities, it underestimates $v_b$ less strongly and $\delta$ more severely.

The new model appears to slightly overestimate the column density and the mass outflow rate at $R_{\rm W}$.  In contrast, the mass outflow rate at $R_{\rm SF}$ is well recovered.  The overestimates of $N_{\rm Si^+}$ and $\dot{M}_{\rm Si^+}(R_{\rm W})$ are likely due to the underestimate of $\delta$ and $\gamma$.  In this context, $\dot{M}_{\rm Si^+}(R_{\rm SF})$ is well recovered because the dependence on $\gamma$ and $\delta$ cancels out.  In equation~\ref{eq:MOR}, the $\gamma-\delta$ dependence suggests that the underestimate of $\delta$ is dominating.  Physically, this would cause SALT to overestimate the amount of mass in the wind at $r>R_{\rm SF}$, as observed.   

Nevertheless, the recovery of ionic column densities and mass-outflow rates from Si II $\lambda\lambda$ 1190,1193 is substantially improved with the new model compared to the older version, particularly at high column densities, $\log(N_{\rm Si^+}/\mathrm{cm}^{-2})\gtrsim15$ (see Figures~17 and 18 of CC23). The remaining biases are smaller than or comparable to the systematic uncertainties reported by \citet{Carr2025_FIRE} from tests of SALT against FIRE-2 simulations: 1.3 dex in $N_{\rm Si^+}$ and $0.36$--$0.63$ dex in mass-outflow rates measured at $0.15$--$0.30\,R_{\rm vir}$. We therefore conclude that the updated SALT model represents a substantial improvement for interpreting metal-ion lines at column densities up to $\log(N_{\rm Si^+}/\mathrm{cm}^{-2})\sim18$, or at comparable line-center optical depths for other transitions. 

Although the additional redward absorption predicted by the updated SALT model appears to improve the recovery of ionic column densities, similar absorption can be produced by the interstellar medium (ISM) or galactic inflows \citep{Carr2022a}, potentially introducing additional degeneracies. Indeed, \citet{Carr2025_FIRE} found that the original SALT model under the Sobolev approximation no longer exhibited a significant bias in the recovered column densities of the FIRE-2 simulations when a Gaussian absorption component centered at the systemic velocity was included. Such a component is commonly used to account for absorption by the ISM \citep{Huberty2024,Carr2025_LyC}. In practice, simultaneously fitting multiple transitions or imposing an independent prior on $v_b$ \citep{Chen2023} may help distinguish these effects and reduce the resulting degeneracies.

\begin{figure*}
	\centering
	\includegraphics[width=\textwidth]{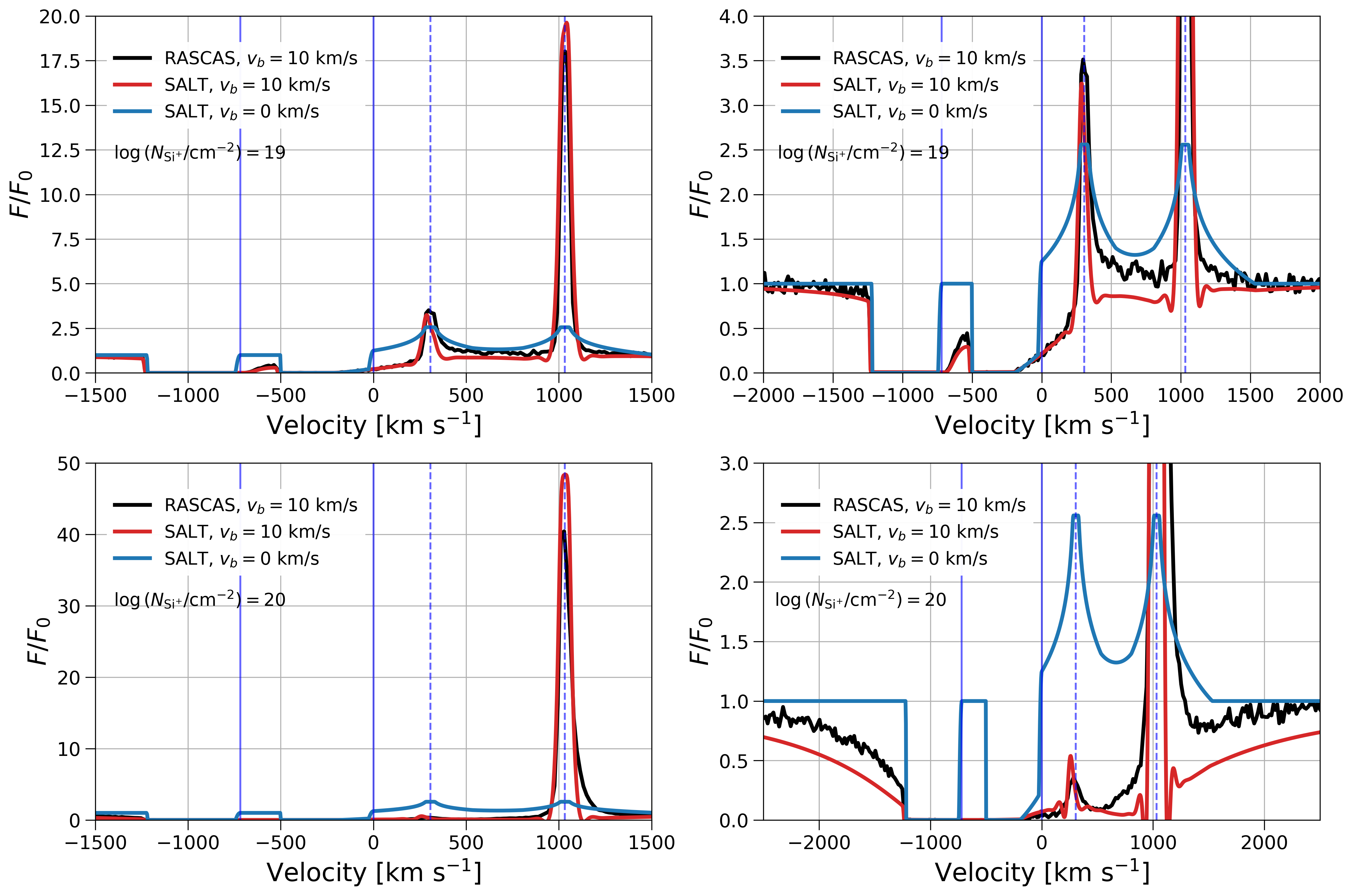}
	\caption{Comparison of RASCAS (red) and SALT (black) predictions for the Si~{\sc ii} $\lambda\lambda1190,1193$ line profiles of spherical $(\alpha = 90^{\circ},\psi = 
    \rm{NA})$ outflows with column densities of $\log(N_{\mathrm{Si}^+}/\mathrm{cm}^{-2})=19$ (top row) and $20$ (bottom row). The light-blue curves show predictions from the original SALT prescription of CC23, which assumes $v_b=0\ \mathrm{km\,s^{-1}}$ and adopts the Sobolev approximation. The left column shows the full line profiles, while the right column provides enlarged views of the absorption features. SALT predicts stronger absorption than RASCAS, with the discrepancy increasing toward higher column densities. At these column densities, the absorption becomes increasingly sensitive to the Lorentzian wings of the cross section.  The number of shells used to compute the emission spectrum is 10.  The rest of the parameters used in these models are $R_{SF} = 1 \ \rm kpc$, $\gamma = 1.0$, $v_b=10\ \mathrm{km\,s^{-1}}$, $v_0 = 25 \ \rm{km \ s^{-1}}$, $v_{\infty} = 500 \ \rm{km \ s^{-1}}$, $v_{ap} = 500 \ \rm{km \ s^{-1}}$, $f_c = 1$, $\delta = 3.0$, and $\kappa = 0$.}
	\label{fig:breakdown}
\end{figure*}

\begin{figure*}
	\centering
	\includegraphics[width=\textwidth]{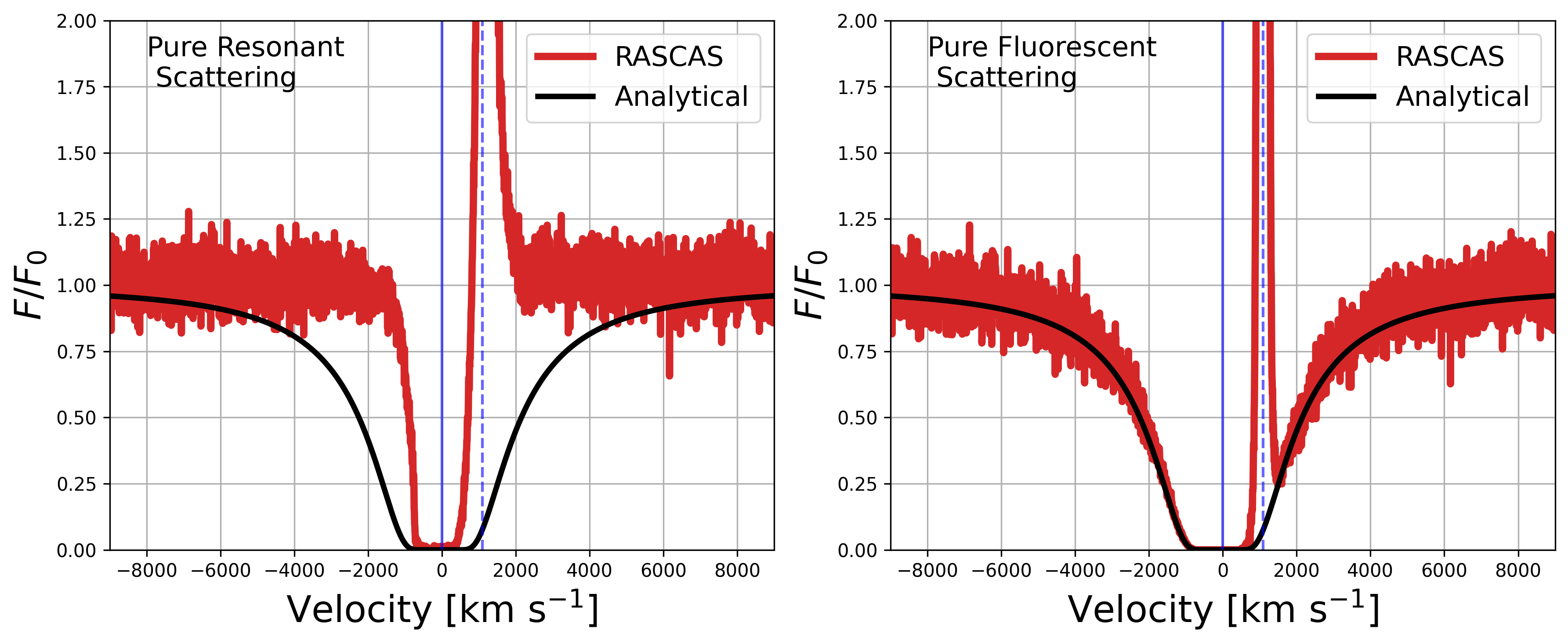}
	\caption{RASCAS predictions (red) for the Si~{\sc ii} \(\lambda1260\) line profiles of spherical ($\alpha=90^{\circ}$,$\psi=\rm NA$) outflows illuminated by a central point source, with \(\log(N_{\mathrm{Si}^+}/\mathrm{cm}^{-2})=20\). The radiative branching probabilities are modified in RASCAS such that absorbed photons are re-emitted exclusively through either the resonant transition at \(1260~\text{\AA}\) (left) or the fluorescent transition at \(1265~\text{\AA}\) (right). The black curves show the analytical pure-absorption solution at resonance.  In the purely resonant case, scattered emission fills in the absorption profile at large observed velocities. In the purely fluorescent case, the RASCAS profile closely follows the pure-absorption solution, indicating little resonant emission infilling. This comparison suggests that resonant frequency diffusion from the line core into the wings of the absorption cross section produces the excess emission at large observed velocities apparent in the discrepancies between RASCAS and SALT in Figure~\ref{fig:breakdown}.  The rest of the parameters used in these models are $R_{SF} = 1 \ \rm kpc$, $\gamma = 1.0$, $v_b=10\ \mathrm{km\,s^{-1}}$, $v_0 = 25 \ \rm{km \ s^{-1}}$, $v_{\infty} = 500 \ \rm{km \ s^{-1}}$, $v_{ap} = 500 \ \rm{km \ s^{-1}}$, $f_c = 1$, $\delta = 3.0$, and $\kappa = 0$.}
	\label{fig:fluor_vs_res}
\end{figure*}

\subsection{Limitations}
We compute the emission line profile in section~\ref{sec:model} assuming that all photons absorbed within a shell are re-emitted from that shell and redistributed in observed velocity space uniformly - albeit for the natural smoothing inherit to thermal and turbulent motion in the gas frame.  In reality, photons may not be re-emitted from the same shell at which they were first absorbed or corresponding frequency.  If the density is high enough, photons can undergo diffusion in both distance and frequency (e.g., \citealt{Smith2018,Lorinc2025}).  We therefore expect our emission line calculations to eventual break down as we continue to increase column density.       

We assess the accuracy of our Si~{\sc ii} $\lambda\lambda1190,1193$ line-profile calculations for spherical outflows with column densities of $\log(N_{\mathrm{Si}^+}/\mathrm{cm}^{-2})=19$ and 20 by comparing them with the RASCAS predictions in the top and bottom rows of Figure~\ref{fig:breakdown}, respectively. These column densities are exceptionally high for $\mathrm{Si}^+$ and are considered solely to examine the behavior and limitations of the model.  The SALT solution predicts much stronger absorption both redward and blueward of line center.  This absorption occurs in the Lorentzian portion of the cross section and is especially visible in the $\log(N_{\mathrm{Si}^+}/\mathrm{cm}^{-2}) = 20$ case.  

After analyzing the RASCAS simulations, we suspect that the discrepancy between SALT and RASCAS is due to diffusion.  We show RASCAS experiments for a single line of Si II $\lambda1260$, generated from a point source of radiation, positioned at the center of a spherical outflow of column density $\log(N_{\mathrm{Si}^+}/\mathrm{cm}^{-2}) = 20$.  This line consists of a resonant transition at 1260\AA\ and a fluorescent line at 1265\AA.  However, we have artificially set the probability of resonant scattering to zero or one, with zero implying a 100\% chance of fluorescent re-emission at 1265\AA\ and one implying no fluorescent re-emission.  

Figure~\ref{fig:fluor_vs_res} shows the purely resonant case in the left panel and the purely fluorescent case in the right panel. For reference, we also show the analytical absorption-only profiles obtained by solving Equation~\ref{eq:absorption_profile} along a single line of sight through the outflow, \(F=F_ce^{-\tau}\). In the purely resonant case, scattered re-emission completely fills in the absorption profile at large observed velocities, corresponding to the Lorentzian component of the cross section, whereas the line profile saturates near zero observed velocity due to absorption in the core. We attribute this behavior to resonant diffusion. In the densest regions of the outflow, photons scatter repeatedly and diffuse in frequency until they reach the wings of the absorption cross section, where the lower opacity allows them to escape. In the purely fluorescent case, absorbed photons are instead re-emitted at fluorescent wavelengths, increasing their probability of escape. The escape probability remains sufficiently high that the photons undergo little subsequent diffusion, thereby limiting emission infilling at large observed velocities.  The fact that the absorption profile resembles the analytical solution suggests that the absorption profiles are the same between the SALT and RASCAS formalisms. The examples in Figure~\ref{fig:breakdown}, represent in between scenarios. 


These extreme test cases and the recovery tests demonstrate that the new SALT model provides a reasonable approximation to the radiative transfer of metal-ion lines, accurately reproducing the Si~{\sc ii} $\lambda\lambda1190,1193$ profiles at column densities of $\log(N_{\rm Si^+}/\mathrm{cm}^{-2})\lesssim18$. At higher column densities, however, frequency and spatial diffusion becomes increasingly important, causing our treatment of scattered re-emission to break down. Although such conditions are unlikely to occur in the metal-ion lines of star-forming galaxies, they are commonly encountered in Ly$\alpha$ radiative transfer, which can probe gas at much higher column densities (e.g., \citealt{Hu2023}).


\section{Conclusion} 
In this work, we present a new version of the Semi-Analytical Line Transfer (SALT) model, a forward model that predicts the spectral-line signatures of galactic winds. The new model accounts for thermal and turbulent motions in the gas frame of the wind. We solve the radiative transfer equation for single-scattering, relaxing the Sobolev approximation where necessary, and validate the model by comparing its spectral predictions with those from Monte Carlo radiative transfer simulations.

We evaluate the recovery of the SALT parameter space by fitting mock spectra generated with Monte Carlo radiative transfer simulations of identical wind configurations using Bayesian Markov chain Monte Carlo sampling. We identify biases in the recovered Doppler parameter ($v_b$) and in the power-law indices of the velocity and density fields. These biases indicate a degeneracy between thermal or turbulent line broadening and the radial structure of the wind: strongly turbulent winds can resemble slowly accelerating winds with shallow density gradients. Nevertheless, the new SALT model substantially improves the recovery of ionic column densities and mass-outflow rates. Although some biases remain, they are smaller than the expected systematic uncertainties inferred from previous recovery tests against hydrodynamic simulations \citep{Carr2025_FIRE}. On average, the ionic column density and mass-outflow rate are overestimated by $0.22^{+0.12}_{-0.14}$ dex and $0.51^{+0.49}_{-0.39}$ dex, respectively.

We also investigate the limitations of the model using synthetic metal-ion lines with exceptionally high column densities, $\log(N_{\rm Si^+}/\mathrm{cm}^{-2})\gtrsim18$. At these column densities, the scattered-emission prescription begins to break down because photons diffuse through both frequency and physical space. Future work will extend the model to transitions that commonly probe this high-optical-depth regime, most notably Ly$\alpha$.

Overall, our recovery tests suggest that global wind properties, including ionic column densities and mass-outflow rates, can be inferred reliably from spatially integrated spectra. By contrast, the radial structures of the velocity and density fields remain difficult to constrain. Spatially resolved integral-field spectroscopy, which maps emission- and absorption-line profiles as functions of projected radius, offers a promising means of breaking these degeneracies. Integral-field spectrographs on the Extremely Large Telescope and, in the longer term, the Habitable Worlds Observatory (HWO) will provide valuable opportunities to constrain radial gradients in galactic outflows \citep{Carr2026}.

\begin{acknowledgments}
C. C. is supported in part by the NSFC grant W2433001 and the NSFC Talent-Introduction Program. C. C. also acknowledges support from the University of Michigan through the ELT Fellowship Program. R. C. acknowledges in part financial support from the start-up funding of Zhejiang University and Zhejiang provincial top level research support program. ChatGPT (OpenAI, GPT-5.2) was used for code optimization and improving clarity of the manuscript text. 
\end{acknowledgments}

%






\appendix

\begin{figure*}
	\centering
	\includegraphics[width=\textwidth]{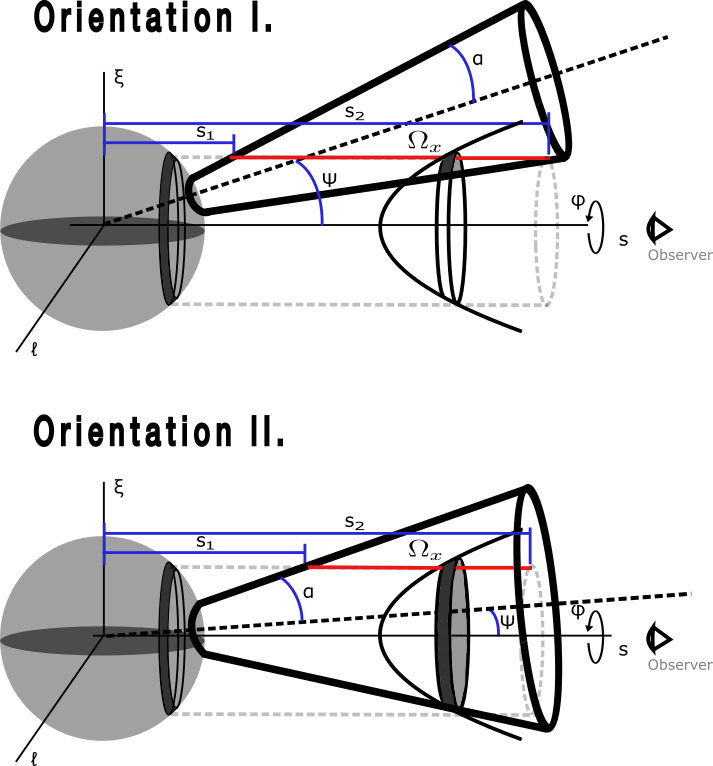}
	\caption{Schematics of bi-conical outflows with opening angle $\alpha$ and orientation angle $\psi$. A surface of constant observed velocity, $\Omega_x$, for a spherical outflow is shown for reference, together with a contour at an arbitrary height (see Figure~\ref{fig:absorption}).  Extending this contour along the $s$-axis defines a cylinder (dashed light grey), whose surface is parameterized by the azimuthal angle $\varphi$ about the $s$-axis. To calculate the absorption profile, we identify the intersections of this cylinder with the bi-cone at each $\varphi$. For fixed $\varphi$, the intersecting path through the outflow (red) extends from $S_1$ to $S_2$. When $\alpha<\psi$ (top; Orientation~I), the path intersects the surface of the cone at both $S_1$ and $S_2$. Otherwise (bottom; Orientation~II), the path enters through the opening of the cone. If, in the latter configuration, $\psi+\alpha>\pi/2$, the path may also intersect the far side of the bi-cone (not shown). The calculation of $S_1$ and $S_2$ at fixed $\varphi$ for each orientation is described in the text.}
	\label{fig:abs_geom}
\end{figure*} 

\section{Bi-conical Outflow}\label{sec:appendix_A}

Here we compute the absorption line profile for a turbulent bi-conical outflow characterized by a half-opening angle, $\alpha$, and an inclination angle, $\psi$, subtended by the axis of the cones and the line of sight.  In the non-turbulent case, under the Sobolev approximation, \cite{Carr2018} found a simple solution by rescaling the line profile for a spherical outflow by a scale factor, $f_g$.  To calculate $f_g$, they decomposed the surface of constant observed velocity $\Omega_x$ into rings centered on the $s$-axis.  See Figure~\ref{fig:abs_geom}.  For a given ring at position, $s$, along the $s$-axis with radius, $h$, they defined $f_g(s,h)$ to be the fraction of the ring intersecting the bi-conical outflow.  Since we are considering turbulent winds, the absorption region occurs over a three-dimensional volume, and the ring calculation must now be performed on a cylinder.  Furthermore, we no longer have azimuthal symmetry around the $s$-axis.  As such, we introduce the azimuthal angle $\varphi=\arctan(l/\xi)$ about the $s$-axis, where $l$ is the Cartesian coordinate perpendicular to both the $s$- and $\xi$-axes. The corresponding azimuthal unit vector is
\begin{equation}
\hat{\boldsymbol{\varphi}}
=
-\sin\varphi\,\hat{\boldsymbol{\xi}}
+\cos\varphi\,\hat{\boldsymbol{l}}.
\end{equation}  In this setting, the absorption line profile for a turbulent bi-conical outflow becomes  
\begin{eqnarray}
    I(x)_{\rm abs, blue}/I_0 &=& \frac{F_{\lambda}}{F_{c,\lambda}} - \frac{1}{\pi}\frac{F_{\lambda}}{F_{c,\lambda}} \int_{0}^{1}\int_0^{2\pi}\nonumber\\ 
    &\times& h \left(1-e^{-\tau(h,\varphi^{\prime})}\right)d\varphi^{\prime}
    dh, \label{eq:absorption_profile_bicone_appendix}
\end{eqnarray}
where the optical depth,
\begin{eqnarray}
    \tau(\varphi,h) = \int_{S_{1}}^{S_{2}}n(s^{\prime})\sigma(s^{\prime})ds^{\prime},
    \label{eq:tau_bicone_appendix}
\end{eqnarray}
is now a function of $\varphi$ and $h$.  $S_{2} - S_{1}$ defines the path through the bi-cone intercepted by the line of sight at a fixed value for $\varphi$ and $h$ (Figure~\ref{fig:abs_geom}, red line). 

To determine the integration limits, $S_1$ and $S_2$, we determine where the line of sight,
\begin{equation}
\vec{r}_{\rm LOS}=\langle s,h\cos\varphi,h\sin\varphi\rangle,
\end{equation}
intersects the boundary of the cone, whose axis is described by the unit vector, 
\begin{equation}
\vec{r}_{\rm BC}=\langle\cos\psi,0,\sin\psi\rangle.
\end{equation}
Following the convention of \cite{Huberty2024}, the line of sight intersects the boundary of the cone when the angle between $\vec{r}_{\rm LOS}$ and $\vec{r}_{\rm BC}$ equals the cone half-opening angle, $\alpha$. Equivalently,
\begin{equation}
\vec{r}_{\rm LOS}\cdot\vec{r}_{\rm BC}
=
|\vec{r}_{\rm LOS}|\cos\alpha.
\end{equation}
This condition reduces to the quadratic equation,
\begin{equation}
As^2+Bs+C=0,
\end{equation}
whose solutions are
\begin{equation}
S_{\pm}
=
\frac{-B\pm\sqrt{B^2-4AC}}{2A},
\end{equation}
where
\begin{equation}
\begin{aligned}
A &= \cos^2\psi - \cos^2\alpha, \\
B &= 2h\sin\varphi\,\cos\psi\,\sin\psi, \\
C &= h^2\left[
      \sin^2\varphi\left(\sin^2\psi-\cos^2\alpha\right)
      -\cos^2\varphi\cos^2\alpha
    \right].
\end{aligned}
\end{equation}

If $\psi > \alpha$ (Orientation I., Figure~\ref{fig:abs_geom}), a line of sight intersects the boundary of the bi-conical outflow at two points (except in the tangent case), yielding
\begin{align}
    S_1 &= \max\!\left(\min(S_{\pm}),S_{\rm SF}\right)\text{ and }
    S_2 = \min\!\left(\max(S_{\pm}),S_{\rm W}\right).
\end{align}
where $S_{\rm SF}$ and $S_{\rm W}$ are the spherical solutions defined in Section~\ref{sec:model}. These limits ensure that the integration neither enters the source region nor extends beyond the outer boundary of the wind. Since only foreground material contributes to the absorption, the integration is restricted to the interval between $S_{\rm SF}$ and $S_{\rm W}$, regardless of whether the intersections arise from the upper or lower cone.

If $\psi \leq \alpha$ (Orientation II., Figure~\ref{fig:abs_geom}), the line of sight lies within the projected opening of the upper cone. In this configuration, the foreground sight line may pass through both the lower and upper cones, with the two absorbing regions separated by the interval between the quadratic roots. Ordering the roots such that
\begin{equation}
    S_- = \min(S_{\pm})
    \text{ and }
    S_+ = \max(S_{\pm}),
\end{equation}
the portion of the sight line within the lower cone is bounded by
\begin{equation}
    S_{1,\mathrm{low}} = S_{\rm SF} \text{ and }
    S_{2,\mathrm{low}} = \min\!\left(S_-,S_{\rm W}\right),
\end{equation}
provided that $S_->S_{\rm SF}$. The portion within the upper cone is bounded by
\begin{equation}
    S_{1,\mathrm{up}} = \max\!\left(S_+,S_{\rm SF}\right) \text{ and }
    S_{2,\mathrm{up}} = S_{\rm W},
\end{equation}
provided that $S_+<S_{\rm W}$. Thus, the optical depth along the sight line is the sum of the contributions from the two disjoint intervals,
\begin{equation}
    \tau =
    \int_{S_{1,\mathrm{low}}}^{S_{2,\mathrm{low}}}
    n(s^{\prime},h) \sigma(\nu_{s^{\prime}}) ds^{\prime}
    +
    \int_{S_{1,\mathrm{up}}}^{S_{2,\mathrm{up}}}
    n(s^{\prime},h) \sigma(\nu_{s^{\prime}}) ds^{\prime},
\end{equation}
where an integral is set to zero if its upper limit does not exceed its lower limit.

\bibliography{bibliography}{}
\bibliographystyle{aasjournal}



\end{document}